\documentclass[sigconf,nonacm]{acmart}

\usepackage{subfigure}
\usepackage{siunitx}
\usepackage{enumitem}

\usepackage{algorithm}
\usepackage[noend]{algpseudocode}
\let\ReturnInline\Return
\renewcommand{\Return}{\State\ReturnInline}
\algrenewcommand\algorithmicrequire{$\rhd$}
\algrenewcommand\algorithmicensure{$\square$}

\AtBeginDocument{%
  \providecommand\BibTeX{{%
    \normalfont B\kern-0.5em{\scshape i\kern-0.25em b}\kern-0.8em\TeX}}}

\setcopyright{acmcopyright}
\copyrightyear{2018}
\acmYear{2018}
\acmDOI{XXXXXXX.XXXXXXX}

\acmConference[Conference acronym 'XX]{Make sure to enter the correct
  conference title from your rights confirmation emai}{June 03--05,
  2018}{Woodstock, NY}

\newcommand{\ignore}[1]{}

\begin{document}

\title[GVE-Leiden: Fast Leiden Algorithm for Community Detection in Shared Memory Setting]{GVE-Leiden: Fast Leiden Algorithm for \\Community Detection in Shared Memory Setting}


\author{Subhajit Sahu}
\email{subhajit.sahu@research.iiit.ac.in}
\affiliation{%
  \institution{IIIT Hyderabad}
  \streetaddress{Professor CR Rao Rd, Gachibowli}
  \city{Hyderabad}
  \state{Telangana}
  \country{India}
  \postcode{500032}
}


\settopmatter{printfolios=true}

\begin{abstract}
Community detection is the problem of identifying natural divisions in networks. Efficient parallel algorithms for identifying such divisions is critical in a number of applications, where the size of datasets have reached significant scales. This technical report presents one of the most efficient implementations of the Leiden algorithm, a high quality community detection method. On a server equipped with dual 16-core Intel Xeon Gold 6226R processors, our Leiden implementation, which we term as GVE-Leiden, outperforms the original Leiden, igraph Leiden, NetworKit Leiden, and cuGraph Leiden (running on NVIDIA A100 GPU) by $436\times$, $104\times$, $8.2\times$, and $3.0\times$ respectively --- achieving a processing rate of $403 M$ edges/s on a $3.8 B$ edge graph. In addition, GVE-Leiden improves performance at an average rate of $1.6\times$ for every doubling of threads.
\end{abstract}



\maketitle

\section{Introduction}
\label{sec:introduction}
Community detection is the problem of identifying subsets of vertices that exhibit higher connectivity among themselves than with the rest of the network. The identified communities are intrinsic when based on network topology alone, and are disjoint when each vertex belongs to only one community. These communities, also known as clusters, shed light on the organization and functionality of the network. This problem with applications in topic discovery, protein annotation, recommendation systems, and targeted advertising \cite{com-gregory10}. One of the difficulties in the community detection problem is the lack of apriori knowledge on the number and size distribution of communities \cite{com-blondel08}. The Louvain method \cite{com-blondel08} is a popular heuristic-based approach for community detection. It employs a two-phase approach, comprising a local-moving phase and an aggregation phase, to iteratively optimize the modularity metric --- a measure of community quality \cite{com-newman06}.

Despite its popularity, the Louvain method has been observed to produce internally-disconnected and badly connected communities. To address these shortcomings, Traag et al. \cite{com-traag19} propose the Leiden algorithm. It introduces an additional refinement phase between the local-moving and aggregation phases. The refinement phase allows vertices to explore and potentially form sub-communities within the communities identified during the local-moving phase. This enables the Leiden algorithm to identify well-connected communities \cite{com-traag19}.

However, applying the original Leiden algorithm to massive graphs has raised computational bottlenecks, mainly due to its inherently sequential nature --- similar to the Louvain method \cite{com-halappanavar17}. In contexts where scalability is paramount, the development of an optimized parallel Leiden algorithm becomes imperative --- especially in the multicore/shared memory setting, due to its energy efficiency and the prevalence of hardware with large memory sizes.\ignore{This technical report addresses precisely this challenge by presenting a parallel implementation of the Leiden algorithm, targeting both quality and efficiency.} Existing studies on parallel Leiden algorithm \cite{verweijfaster, nguyenleiden} propose a number of parallelization techniques, but do not address optimization for the aggregation phase of the Leiden algorithm, which emerges as a bottleneck after the local-moving phase of the algorithm has been optimized. In addition, a number of optimization techniques that apply to the Louvain method also apply to the Leiden algorithm.\ignore{However, these are scattered over a number of papers, making it difficult for a reader to get a grip over them.}

In this report, we present our parallel multicore implementation of the Leiden algorithm\footnote{\url{https://github.com/puzzlef/leiden-communities-openmp}}. It incorporates several optimizations, including parallel prefix sums, preallocated Compressed Sparse Row (CSR) data structures for community vertex identification and super-vertex graph storage during aggregation, fast collision-free per-thread hash tables for the local-moving and aggregation phases, and prevention of unnecessary aggregations --- enabling it to run significantly faster than existing implementations. Additionally, we employ a greedy refinement phase where vertices optimize for delta-modularity within their community bounds, yielding improved performance and quality compared to a randomized approach. Furthermore, we utilize established techniques such as asynchronous computation, OpenMP's dynamic loop schedule, threshold-scaling optimization, and vertex pruning. To the best of our knowledge, our implementation is the most efficient implementation of Leiden algorithm on multicore CPUs to date. We conduct comprehensive comparisons with other state-of-the-art implementations, including multi-core, GPU-based implementations, multi-node implementations, detailed in Table \ref{tab:compare}. Both direct and indirect comparisons are provided, with details given in Sections \ref{sec:comparison} and \ref{sec:comparison-indirect}, respectively.

\begin{table}[hbtp]
  \centering
  \caption{Speedup of our multicore implementation of Leiden algorithm compared to other state-of-the-art implementations. Direct comparisons entail running the given implementation on our servers, while indirect comparisons (marked with a $*$, explained in Section \ref{sec:comparison-indirect}) involve comparing results relative to a common reference\ignore{(original Leiden)}.\ignore{Notably, the Leiden implementations vary in their classification, with some being multi-core and others multi-node.}}
  \label{tab:compare}
  \begin{tabular}{|c|c||c|}
    \toprule
    \textbf{Leiden implementation} &
    \textbf{Parallelism} &
    \textbf{Our Speedup} \\
    \midrule
    Original Leiden \cite{com-traag19} & Sequential & $436\times$ \\ \hline
    igraph Leiden \cite{csardi2006igraph} & Sequential & $104\times$ \\ \hline
    NetworKit Leiden \cite{staudt2016networkit} & Parallel & $8.2\times$ \\ \hline
    cuGraph Leiden \cite{kang2023cugraph} & Parallel (GPU) & $3.0\times$ \\ \hline
    ParLeiden-S \cite{huparleiden} & Parallel & $18\times^*$ \\ \hline
    ParLeiden-D (8 nodes) \cite{huparleiden} & Multi-node & $22\times^*$ \\ \hline
  \bottomrule
  \end{tabular}
\end{table}

\section{Related work}
\label{sec:related}
The Louvain method, introduced by Blondel et al. \cite{com-blondel08} from the University of Louvain, is a greedy modularity-optimization based algorithm for community detection \cite{com-lancichinetti09}. While it is favored for identifying communities with high modularity, it often results in internally disconnected communities. This occurs when a vertex, acting as a bridge, moves to another community during iterations. Further iterations aggravate the problem, without decreasing the quality function. Further, the Louvain method may identify communities that are not well connected, i.e., splitting certain communities could improve the quality score --- such as modularity \cite{com-traag19}.\ignore{This is not the same as resolution limit problem with modularity, that causes small communities to be clustered with large communities. Louvain only guarantees that no communities can be merged (well separated).}

To address these limitations, Traag et al. \cite{com-traag19} from the University of Leiden, propose the Leiden algorithm\ignore{as an enhancement of the Louvain method}. It introduces a \textit{refinement phase} after the local-moving phase, where vertices within each community undergo constrained merges in a randomized fashion proportional to the delta-modularity of the move. This allows vertices to find sub-communities within those obtained from the local-moving phase. The Leiden algorithm guarantees that the identified communities are both well separated (like the Louvain method) and well connected. When communities have converged, it is guaranteed that all vertices are optimally assigned, and all communities are subset optimal \cite{com-traag19}. Shi et al. \cite{com-shi21} also introduce an additional refinement phase after the local-moving phase with the Louvain method, which they observe minimizes bad clusters. It should however be noted that methods relying on modularity maximization are known to suffer from resolution limit problem, which prevents identification of communities of certain sizes \cite{com-ghosh19, com-traag19}. This can be overcome by using an alternative quality function, such as the Constant Potts Model (CPM) \cite{com-traag11}.

We now discuss a number of algorithmic improvements proposed for the Louvain method, that also apply to the Leiden algorithm. These include ordering of vertices based on importance \cite{com-aldabobi22}, attempting local move only on likely vertices \cite{com-ryu16, com-zhang21, com-shi21}\ignore{\cite{com-ozaki16}}, early pruning of non-promising candidates\ignore{(leaf vertices)} \cite{com-ryu16, com-halappanavar17, com-zhang21, com-you22}, moving vertices to a random neighbor community \cite{com-traag15}, subnetwork refinement \cite{com-waltman13, com-traag19}, multilevel refinement \cite{com-rotta11, com-gach14, com-shi21}, threshold cycling \cite{com-ghosh18}, threshold scaling \cite{com-lu15, com-naim17, com-halappanavar17}, and early termination \cite{com-ghosh18}. A number of parallelization techniques have been also attempted for the Louvain method, that may also be applied to the Leiden algorithm. These include\ignore{using heuristics to break the sequential barrier \cite{com-lu15},} using adaptive parallel thread assignment \cite{com-fazlali17, com-naim17, com-sattar19, com-mohammadi20}, parallelizing the costly first iteration \cite{com-wickramaarachchi14}, ordering vertices via graph coloring \cite{com-halappanavar17}, performing iterations asynchronously \cite{com-shi21}\ignore{\cite{com-que15}}\ignore{, using vector based hashtables \cite{com-halappanavar17}}, and using sort-reduce instead of hashing \cite{com-cheong13}\ignore{, using simple partitions based of vertex ids \cite{com-cheong13, com-ghosh18}, and identifying and moving ghost/doubtful vertices \cite{com-zeng15, com-que15, com-bhowmik19, com-bhowmick22}}.\ignore{Platforms used range from an AMD multicore system \cite{com-fazlali17}, and Intel’s Knight's Landing, Haswell \cite{com-gheibi20}, SkylakeX, and Cascade Lake \cite{part-hossain21}. Other approaches include the use of MapReduce in a BigData batch processing framework \cite{com-zeitz17}.}

A few open source implementations and software packages have been developed for community detection using Leiden algorithm. The original implementation of the Leiden algorithm \cite{com-traag19}, called \texttt{libleidenalg}, is written in C++ and has a Python interface called \texttt{leidenalg}. NetworKit \cite{staudt2016networkit} is a software package designed for analyzing the structural aspects of graph data sets with billions of connections. It utilizes a hybrid with C++ kernels and a Python frontend. The package features a parallel implementation of the Leiden algorithm by Nguyen \cite{nguyenleiden} which uses global queues for vertex pruning, and vertex and community locking for updating communities. igraph \cite{csardi2006igraph} is a similar package, written in C, with Python, R, and Mathematica frontends. It is widely used in academic research, and includes an implementation of the Leiden algorithm. cuGraph \cite{kang2023cugraph} is a GPU-accelerated graph analytics library, part of the RAPIDS suite of data science and machine learning libraries. It leverages the computational power of NVIDIA GPUs to perform graph analytics tasks much faster than traditional CPU-based approaches. cuGraph's core is implemented in C++ with CUDA and is used primarily through its Python interface, making it convenient for data scientists and developers\ignore{working in Python}.

However, most existing works only focus on optimizing the local-moving phase of Leiden algorithm, are not sufficiently optimized, and are either sequential or lack effective parallelization. For instance, the \textit{original Leiden} implementation \cite{com-traag19}\ignore{, \textit{libleidenalg}}, uses a full \texttt{Graph} data structure for each of the collapsed graphs without reuse and makes a complete copy of the communities when collapsing the graph. Further, it tracks the community membership of vertices in each dendrogram layer, adding to the computational overhead. Additionally, it employs a queue and a flag vector to monitor the nodes to be moved in each iteration, with the processing order randomly shuffled. A new degree vector is allocated in each iteration, and community memberships are renumbered after each local-moving and refinement phase. This implementation also conducts several book-keeping activities when moving a node to another community, such as maintaining community sets, removing empty communities, and updating the total edge weight from communities. Finally, it also  keeps track of nodes with fixed community memberships using three vectors, a process repeated in each local-moving phase.

The \textit{igraph Leiden} implementation \cite{csardi2006igraph} follows a comparable approach but with some differences that streamline its operations slightly. Like the original Leiden, it uses a full \texttt{Graph} data structure for each collapsed graph without reuse. igraph Leiden also uses a queue and a flag vector to manage nodes during the local-moving phase, with the processing order randomly shuffled as well. In addition, it also reindexes the community membership of vertices after the local-moving phase, and the refinement phase also involves a random node shuffling. However, unlike the original Leiden, igraph only maintains the size of each community rather than the individual nodes, simplifying the process. Unlike the original Leiden, it does not track edge weights from communities but instead keeps track of total edge weights, which is easier to manage.

The \textit{NetworKit Leiden} implementation \cite{staudt2016networkit} introduces parallel processing but retains some of the complexity seen in the above implementations. It uses a queue and a flag vector to track vertices for processing, incorporating a mutex and a condition variable for parallel access and updates, with node shuffling done on a single thread. Synchronization between threads is performed to manage available work, while employing locks. During the aggregation phase, not all operations are parallelized, such as the prefix sum calculation. Community membership of vertices is tracked and flattened at the end, while node mapping is stored in a \texttt{std::map}, and fine to coarse mapping is sequentially stored. These issues\ignore{, along with non-parallelized aggregation tasks,} contribute to the computational expense of NetworKit Leiden, despite the benefits of parallel processing.

\section{Preliminaries}
\label{sec:preliminaries}
Consider an undirected graph $G(V, E, w)$, where $V$ represents the set of vertices, $E$ the set of edges, and $w_{ij} = w_{ji}$ denotes the weight associated with each edge. In the case of an unweighted graph, we assume unit weight for each edge ($w_{ij} = 1$). Additionally, the neighbors of a vertex $i$ are denoted as $J_i = \{j\ |\ (i, j) \in E\}$, the weighted degree of each vertex as $K_i = \sum_{j \in J_i} w_{ij}$, the total number of vertices as $N = |V|$, the total number of edges as $M = |E|$, and the sum of edge weights in the undirected graph as $m = \sum_{i, j \in V} w_{ij}/2$.

\subsection{Community detection}

Disjoint community detection involves identifying a community membership mapping, $C: V \rightarrow \Gamma$, where each vertex $i \in V$ is assigned a community-id $c$ from the set of community-ids $\Gamma$. We denote the vertices of a community $c \in \Gamma$ as $V_c$, and the community that a vertex $i$ belongs to as $C_i$. Further, we denote the neighbors of vertex $i$ belonging to a community $c$ as $J_{i \rightarrow c} = \{j\ |\ j \in J_i\ and\ C_j = c\}$, the sum of those edge weights as $K_{i \rightarrow c} = \sum_{j \in J_{i \rightarrow c}} w_{ij}$, the sum of weights of edges within a community $c$ as $\sigma_c = \sum_{(i, j) \in E\ and\ C_i = C_j = c} w_{ij}$, and the total edge weight of a community $c$ as $\Sigma_c = \sum_{(i, j) \in E\ and\ C_i = c} w_{ij}$ \cite{com-leskovec21}.

\subsection{Modularity}

Modularity serves as a\ignore{fitness} metric for evaluating the quality of communities identified by heuristic-based community detection algorithms. It is calculated as the difference between the fraction of edges within communities and the expected fraction if edges were randomly distributed, yielding a range of $[-0.5, 1]$, where higher values signify superior results \cite{com-brandes07}.\ignore{The optimization of this metric theoretically leads to the optimal grouping \cite{com-newman04, com-traag11}.} The modularity $Q$ of identified communities is determined using Equation \ref{eq:modularity}, where $\delta$ represents the Kronecker delta function ($\delta (x,y)=1$ if $x=y$, $0$ otherwise). The \textit{delta modularity} of moving a vertex $i$ from community $d$ to community $c$, denoted as $\Delta Q_{i: d \rightarrow c}$, can be calculated using Equation \ref{eq:delta-modularity}.

\begin{equation}
\label{eq:modularity}
  Q
  = \frac{1}{2m} \sum_{(i, j) \in E} \left[w_{ij} - \frac{K_i K_j}{2m}\right] \delta(C_i, C_j)
  = \sum_{c \in \Gamma} \left[\frac{\sigma_c}{2m} - \left(\frac{\Sigma_c}{2m}\right)^2\right]
\end{equation}

\begin{equation}
\label{eq:delta-modularity}
  \Delta Q_{i: d \rightarrow c}
  = \frac{1}{m} (K_{i \rightarrow c} - K_{i \rightarrow d}) - \frac{K_i}{2m^2} (K_i + \Sigma_c - \Sigma_d)
\end{equation}

\subsection{Louvain algorithm}
\label{sec:about-louvain}

The Louvain method \cite{com-blondel08} is an agglomerative algorithm that optimizes modularity to identify high quality disjoint communities in large networks. It has a time complexity of $O (L |E|)$, where $L$ is the total number of iterations performed, and a space complexity of $O(|V| + |E|)$ \cite{com-lancichinetti09}. This algorithm comprises two phases: the \textit{local-moving phase}, in which each vertex $i$ greedily decides to join the community of one of its neighbors $j \in J_i$ to maximize the increase in modularity $\Delta Q_{i:C_i \rightarrow C_j}$ (using Equation \ref{eq:delta-modularity}), and the \textit{aggregation phase}, where all vertices in a community are merged into a single super-vertex. These phases constitute one pass, which is repeated until there is no further increase in modularity is observed \cite{com-blondel08, com-leskovec21}.\ignore{We observe that Louvain obtains high-quality communities, with $3.0 - 30\%$ higher modularity than that obtained by LPA, but requires $2.3 - 14\times$ longer to converge.}

\subsection{Leiden algorithm}
\label{sec:about-leiden}

As mentioned earlier, the Louvain method, while effective, may identify internally disconnected communities and arbitrarily badly connected ones. Traag et al. \cite{com-traag19} proposed the Leiden algorithm to address these issues. The algorithm introduces a \textit{refinement phase} subsequent to the local-moving phase, wherein vertices within each community undergo constrained merges to other sub-communities within their community bounds (obtained from the local-moving phase), starting from a singleton sub-community. This is performed in a randomized manner, with the probability of joining a neighboring sub-community within its community bound being proportional to the delta-modularity of the move. This facilitates the identification of sub-communities within those obtained from the local-moving phase.\ignore{The Leiden algorithm not only guarantees that all communities are well separated (akin to the Louvain method), but also are well connected.} Once communities have converged, it is guaranteed that all vertices are optimally assigned, and all communities are subset optimal \cite{com-traag19}. It has a time complexity of $O (L |E|)$, where $L$ is the total number of iterations performed, and a space complexity of $O(|V| + |E|)$, similar to the Louvain method.

\section{Approach}
\label{sec:approach}
\subsection{Optimizations for Leiden algorithm}
\label{sec:leiden}

We extend optimization techniques, originally designed for the Louvain method \cite{sahu2023gvelouvain}, to the Leiden algorithm. Specifically, we implement an \textit{asynchronous} version of the Leiden algorithm, allowing threads to operate independently on distinct sections of the graph. While this approach promotes faster convergence, it also introduces variability into the final result \cite{com-shi21}. To ensure efficient computations, we allocate a dedicated hashtable per thread. These hashtables serve two main purposes: they keep track of the delta-modularity associated with moving to each community connected to a vertex during the local-moving/refinement phases, and they record the total edge weight between super-vertices in the aggregation phase of the algorithm \cite{sahu2023gvelouvain}.

Our optimizations include using preallocated Compressed Sparse Row (CSR) data structures for identifying community vertices ($G'_{C'}$ in Algorithm \ref{alg:leidenag}) and storing the super-vertex graph ($G''$ in Algorithm \ref{alg:leidenag}) during aggregation, utilizing parallel prefix sum (lines \ref{alg:leidenag--coff-begin}-\ref{alg:leidenag--coff-end}, \ref{alg:leidenag--yoff-end} in Algorithm \ref{alg:leidenag}), employing fast collision-free per-thread hashtables that are well separated in their memory addresses ($H_t$ in Algorithms \ref{alg:leidenlm}, \ref{alg:leidenre}, and \ref{alg:leidenag}), and using an aggregation tolerance of $0.8$ to avoid performing aggregations of minimal utility (line \ref{alg:leiden--aggregation-tolerance} in Algorithm \ref{alg:leidenag}). Additionally, we implement flag-based vertex pruning (lines \ref{alg:leidenlm--reset-affected}, \ref{alg:leidenlm--prune}, \ref{alg:leidenlm--remark} in Algorithm \ref{alg:leidenlm}) --- instead of a queue-based one \cite{nguyenleiden}, utilize OpenMP's \textit{dynamic} loop scheduling, cap the number of iterations per pass at $20$ (line \ref{alg:leidenlm--iterations-begin} in Algorithm \ref{alg:leidenlm}), employ a tolerance drop rate of $10$ (line \ref{alg:leiden--threshold-scaling} in Algorithm \ref{alg:leiden}) --- threshold scaling optimization, and initiate with a tolerance of $0.01$ \cite{sahu2023gvelouvain}.

\begin{figure}[hbtp]
  \centering
  \subfigure{
    \label{fig:leidenopt-runtime--all}
    \includegraphics[width=0.98\linewidth]{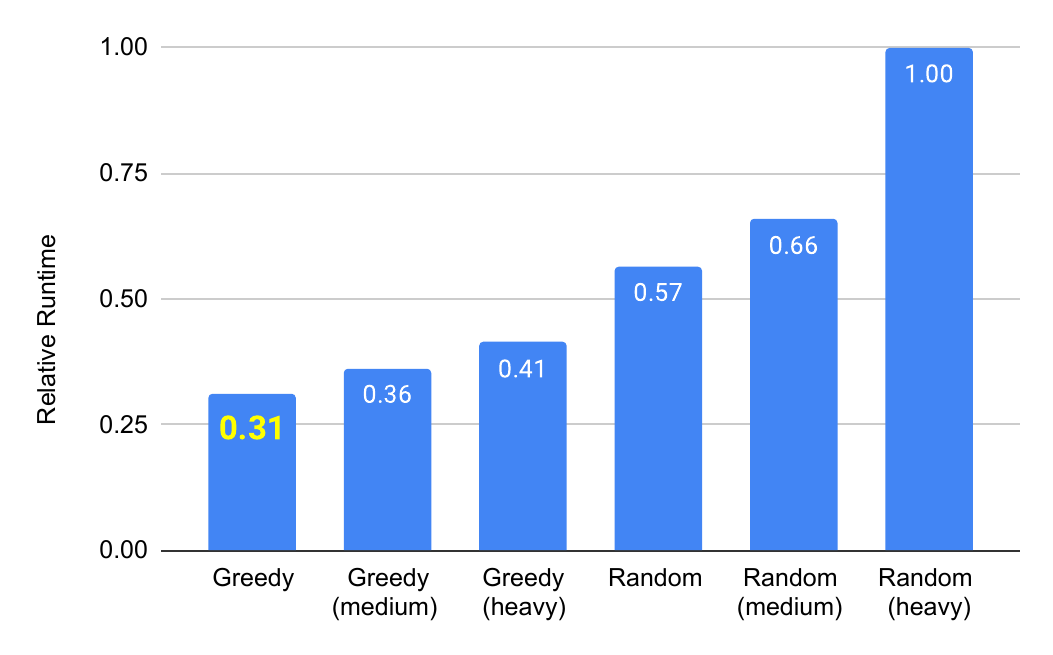}
  } \\[-2ex]
  \caption{Average relative runtime for the \textit{greedy} and \textit{random} approaches (including \textit{medium} and \textit{heavy} variants) of parallel Leiden algorithm for all graphs in the dataset.}
  \label{fig:leidenopt-runtime}
\end{figure}

\begin{figure}[hbtp]
  \centering
  \subfigure{
    \label{fig:leidenopt-modularity--all}
    \includegraphics[width=0.98\linewidth]{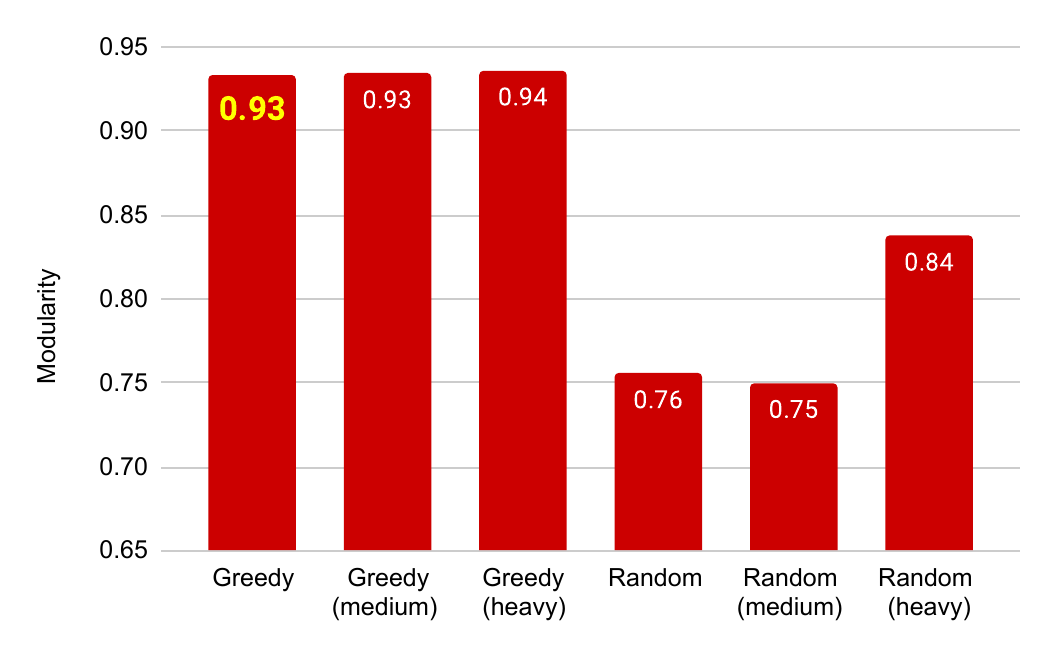}
  } \\[-2ex]
  \caption{Average modularity for the \textit{greedy} and \textit{random} approaches (including \textit{medium} and \textit{heavy} variants) of parallel Leiden algorithm for all graphs in the dataset.}
  \label{fig:leidenopt-modularity}
\end{figure}

We attempt two approaches of the Leiden algorithm. One uses a \textit{greedy refinement phase} where vertices greedily optimize for delta-modularity (within their community bounds), while the other uses a \textit{randomized refinement phase} (using fast \textit{xorshift32} random number generators), where the likelihood of selection of a community to move to (by a vertex) is proportional to its delta-modularity, as originally proposed \cite{com-traag19}. Our results, shown in Figures \ref{fig:leidenopt-runtime} and \ref{fig:leidenopt-modularity}, indicate the \textit{greedy approach} performs the best on average, both in terms of runtime and modularity. We also try medium and heavy variants for both approaches, which disables threshold scaling and aggregation tolerance (including threshold scaling) respectively, However, we do not find them to perform well overall.\ignore{On \textit{europe\_osm} graph, our parallel Greedy-Leiden (which we from here on refer to simply as Leiden) runs $3\times$ faster than Nguyen \cite{nguyenleiden}.}

\ignore{\subsection{Community labels of super-vertices}}

\begin{figure}[hbtp]
  \centering
  \subfigure{
    \label{fig:leidenreopt-runtime--all}
    \includegraphics[width=0.98\linewidth]{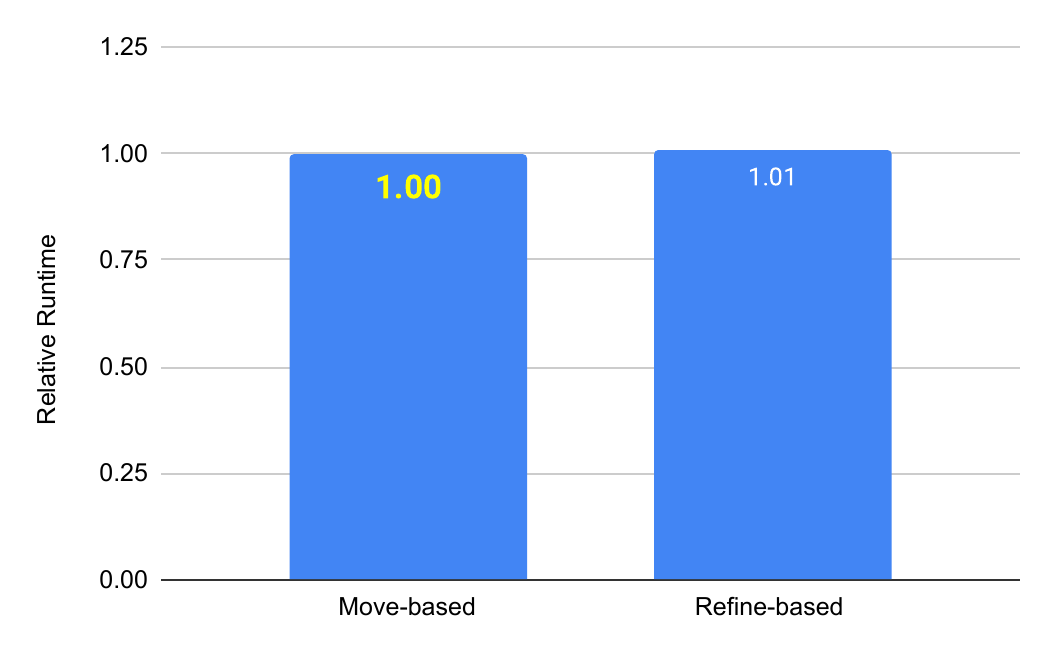}
  } \\[-2ex]
  \caption{Average relative runtime for \textit{move-based} and \textit{refine-based} communities for super-vertices upon aggregation with parallel Leiden algorithm, for all graphs in the dataset.}
  \label{fig:leidenreopt-runtime}
\end{figure}

\begin{figure}[hbtp]
  \centering
  \subfigure{
    \label{fig:leidenreopt-modularity--all}
    \includegraphics[width=0.98\linewidth]{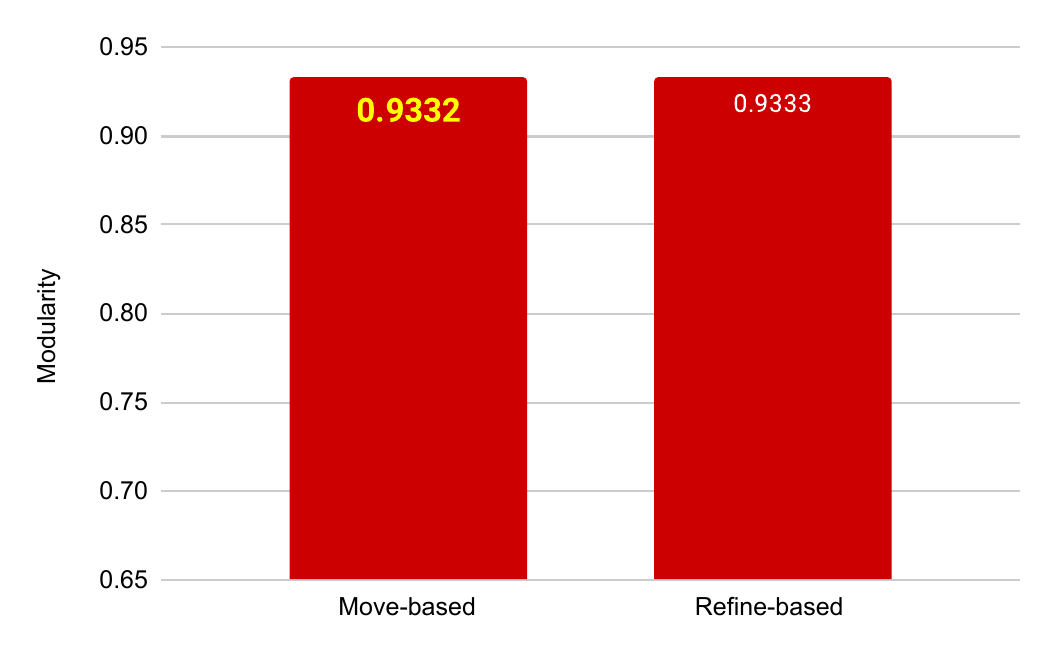}
  } \\[-2ex]
  \caption{Average modularity for \textit{move-based} and \textit{refine-based} communities for super-vertices upon aggregation with parallel Leiden algorithm, for all graphs in the dataset.}
  \label{fig:leidenreopt-modularity}
\end{figure}

We also attempt two different variations of Parallel Leiden algorithm, one where the community labels of super-vertices (upon aggregation) is based on the local-moving phase (\textit{move-based}), and the other where the community labels of super-vertices is based on the refinement phase (\textit{refine-based}). Our observations indicate that both approaches have roughly the same runtime and modularity on average, as indicated by Figures \ref{fig:leidenreopt-runtime} and \ref{fig:leidenreopt-modularity}. Accordingly, we stick to the move-based approach, which is the one recommended by Traag et al. \cite{com-traag19}. However, refine-based approach may be more suitable for the design of dynamic Leiden algorithm (for dynamic graphs).

\ignore{We fixed a bug that caused the Leiden algorithm to fail in finding communities on road networks and kmer graphs. The issue was forgetting to reset the affected vertices flags before running the refinement phase.}

\begin{figure*}[hbtp]
  \centering
  \subfigure{
    \label{fig:leiden-pass--all}
    \includegraphics[width=0.98\linewidth]{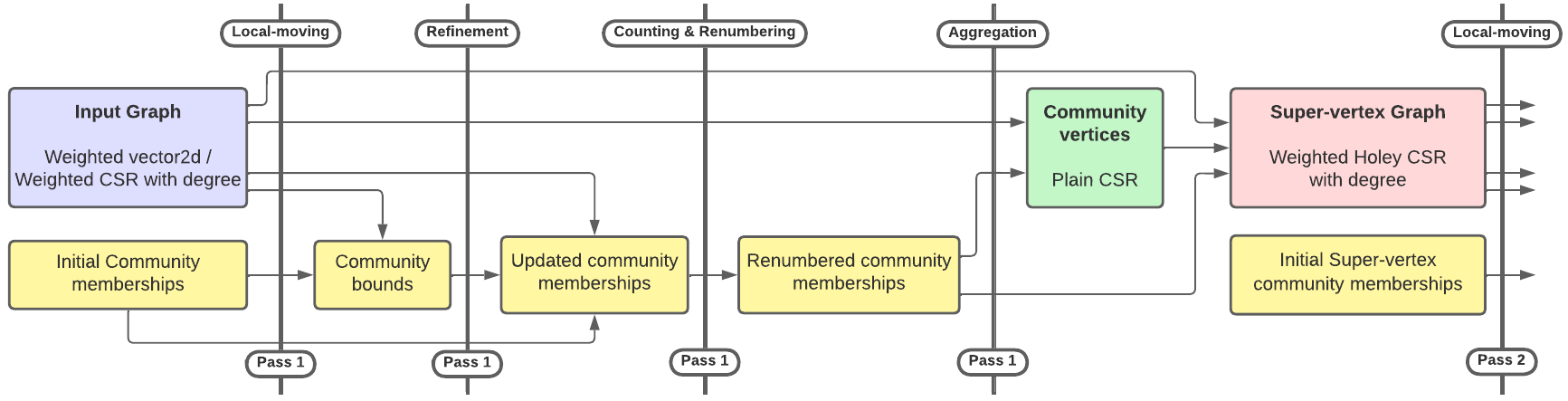}
  } \\[-2ex]
  \caption{A flow diagram illustrating the first pass of GVE-Leiden for either a Weighted 2D-vector-based or a Weighted CSR with degree-based input graph. In the local-moving phase, vertex community memberships are updated to obtain community bounds for the refinement phase, until the cumulative delta-modularity change across all vertices reaches a specified threshold. Then, in the refinement phase, the each vertex starts in a singleton community, and community memberships are updated similarly to the local-moving phase, with vertices changing communities within their bounds. These community memberships are then counted and renumbered. In the aggregation phase, community vertices in a CSR are first obtained. This is used to create the super-vertex graph stored in a Weighted Holey CSR with degree. Subsequent passes use a Weighted Holey CSR with degree and initial community memberships for super-vertices from the previous pass as input.}
  \label{fig:leiden-pass}
\end{figure*}

\subsection{Our optimized Leiden implementation}

We now explain the implementation of GVE-Leiden in Algorithms \ref{alg:leiden}, \ref{alg:leidenlm}, \ref{alg:leidenre}, and \ref{alg:leidenag}. Our intention is to integrate GVE-Leiden into a forthcoming command-line graph processing tool named "GVE", which simply stands for Graph(Vertices, Edges), hence the name. GVE-Leiden operates with a time complexity of $O(KM)$, where $K$ is the total number of iterations performed, and a space complexity of $O(TN + M)$, where $T$ represents the number of threads, and $TN$ accounts for the collision-free hash tables $H_t$ used per thread. Figure \ref{fig:leiden-pass} illustrates the first pass of GVE-Leiden.

\subsubsection{Main step of GVE-Leiden}

The main step of GVE-Leiden (\texttt{leiden()} function) is outlined in Algorithm \ref{alg:leiden}. It encompasses initialization, the local-moving phase, the refinement phase, and the aggregation phase. Here, the \texttt{leiden()} function accepts the input graph $G$, and returns the community membership $C$ of each vertex. In line \ref{alg:leiden--initialization}, we first initialize the community membership $C$ for each vertex in $G$, and perform passes of the Leiden algorithm, limited to $MAX\_PASSES$ (lines \ref{alg:leiden--passes-begin}-\ref{alg:leiden--passes-end}). During each pass, we initialize the total edge weight of each vertex $K'$, the total edge weight of each community $\Sigma'$, and the community membership $C'$ of each vertex in the current graph $G'$ (line \ref{alg:leiden--reset-weights}).

Subsequently, in line \ref{alg:leiden--local-move}, we perform the local-moving phase by invoking \texttt{leidenMove()}\ignore{(Algorithm \ref{alg:leidenlm})}, which optimizes community assignments. Following this, we set the \textit{community bound} of each vertex (for the refinement phase) as the community membership of each vertex just obtained, and reset the membership of each vertex, and the total weight of each community as singleton vertices in line \ref{alg:leiden--reset-again}. In line \ref{alg:leiden--refine}, the refinement phase is carried out by invoking \texttt{leidenRefine()}\ignore{(Algorithm \ref{alg:leidenre})}, which optimizes the community assignment of each vertex within its community bound. If the local-moving phase converges in a single iteration, global convergence is implied and we terminate the passes (line \ref{alg:leiden--globally-converged}). Further, if the drop in the number of communities $|\Gamma|$ is marginal, we halt the algorithm at the current pass (line \ref{alg:leiden--aggregation-tolerance}).

If convergence has not been achieved, we proceed to renumber communities (line \ref{alg:leiden--renumber}), update top-level community memberships $C$ with dendrogram lookup (line \ref{alg:leiden--lookup}), perform the aggregation phase by calling \texttt{leidenAggregate()}\ignore{(Algorithm \ref{alg:leidenag})}, and adjust the convergence threshold for subsequent passes, i.e., perform threshold scaling (line \ref{alg:leiden--threshold-scaling}). The next pass commences in line \ref{alg:leiden--passes-begin}. At the end of all passes, we perform a final update of the top-level community memberships $C$ with dendrogram lookup (line \ref{alg:leiden--lookup-last}), and return the top-level community membership $C$ of each vertex in $G$.

\begin{algorithm}[hbtp]
\caption{GVE-Leiden: Our parallel Leiden algorithm.}
\label{alg:leiden}
\begin{algorithmic}[1]
\Require{$G$: Input graph}
\Require{$C$: Community membership of each vertex}
\Require{$G'$: Input/super-vertex graph}
\Require{$C'$: Community membership of each vertex in $G'$}
\Require{$K'$: Total edge weight of each vertex}
\Require{$\Sigma'$: Total edge weight of each community}
\Ensure{$l_i$, $l_j$: Number of iterations performed (per pass)}
\Ensure{$l_p$: Number of passes performed}
\Ensure{$\tau$: Per iteration tolerance}
\Ensure{$\tau_{agg}$: Aggregation tolerance}

\Statex

\Function{leiden}{$G$} \label{alg:leiden--begin}
  \State Vertex membership: $C \gets [0 .. |V|)$ \textbf{;} $G' \gets G$ \label{alg:leiden--initialization}
  \ForAll{$l_p \in [0 .. \text{\small{MAX\_PASSES}})$} \label{alg:leiden--passes-begin}
    \State $\Sigma' \gets K' \gets vertexWeights(G')$ \textbf{;} $C' \gets [0 .. |V'|)$ \label{alg:leiden--reset-weights}
    \State $l_i \gets leidenMove(G', C', K', \Sigma', \tau)$ \Comment{Alg. \ref{alg:leidenlm}} \label{alg:leiden--local-move}
    \State $C'_B \gets C'$ \textbf{;} $C' \gets [0 .. |V'|)$ \textbf{;} $\Sigma' \gets K'$ \label{alg:leiden--reset-again}
    \State $l_j \gets leidenRefine(G', C'_B, C', K', \Sigma')$ \Comment{Alg. \ref{alg:leidenre}} \label{alg:leiden--refine}
    \If{$l_i + l_j \le 1$} \textbf{break} \Comment{Globally converged?} \label{alg:leiden--globally-converged}
    \EndIf
    \State $|\Gamma|, |\Gamma_{old}| \gets$ Number of communities in $C$, $C'$
    \If{$|\Gamma|/|\Gamma_{old}| > \tau_{agg}$} \textbf{break} \Comment{Low shrink?} \label{alg:leiden--aggregation-tolerance}
    \EndIf
    \State $C' \gets$ Renumber communities in $C'$ \label{alg:leiden--renumber}
    \State $C \gets$ Lookup dendrogram using $C$ to $C'$ \label{alg:leiden--lookup}
    \State $G' \gets leidenAggregate(G', C')$ \Comment{Alg. \ref{alg:leidenag}} \label{alg:leiden--aggregate}
    \State $C' \gets$ Map $C'$ to $C'_B$ \Comment{Use move-based membership} \label{alg:leiden--useparent}
    \State $\tau \gets \tau / \text{\small{TOLERANCE\_DROP}}$ \Comment{Threshold scaling} \label{alg:leiden--threshold-scaling}
  \EndFor \label{alg:leiden--passes-end}
  \State $C \gets$ Lookup dendrogram using $C$ to $C'$ \label{alg:leiden--lookup-last}
  \Return{$C$} \label{alg:leiden--return}
\EndFunction \label{alg:leiden--end}
\end{algorithmic}
\end{algorithm}

\begin{algorithm}[hbtp]
\caption{Local-moving phase of GVE-Leiden.}
\label{alg:leidenlm}
\begin{algorithmic}[1]
\Require{$G'$: Input/super-vertex graph}
\Require{$C'$: Community membership of each vertex}
\Require{$K'$: Total edge weight of each vertex}
\Require{$\Sigma'$: Total edge weight of each community}
\Ensure{$H_t$: Collision-free per-thread hashtable}
\Ensure{$l_i$: Number of iterations performed}
\Ensure{$\tau$: Per iteration tolerance}

\Statex

\Function{leidenMove}{$G', C', K', \Sigma', \tau$} \label{alg:leidenlm--move-begin}
  \State Mark all vertices in $G'$ as unprocessed \label{alg:leidenlm--reset-affected}
  \ForAll{$l_i \in [0 .. \text{\small{MAX\_ITERATIONS}})$} \label{alg:leidenlm--iterations-begin}
    \State Total delta-modularity per iteration: $\Delta Q \gets 0$ \label{alg:leidenlm--init-deltaq}
    \ForAll{unprocessed $i \in V'$ \textbf{in parallel}} \label{alg:leidenlm--loop-vertices-begin}
      \State Mark $i$ as processed (prune) \label{alg:leidenlm--prune}
      \State $H_t \gets scanCommunities(\{\}, G', C', i, false)$ \label{alg:leidenlm--scan}
      \State $\rhd$ Use $H_t, K', \Sigma'$ to choose best community
      \State $c^* \gets$ Best community linked to $i$ in $G'$ \label{alg:leidenlm--best-community-begin}
      \State $\delta Q^* \gets$ Delta-modularity of moving $i$ to $c^*$ \label{alg:leidenlm--best-community-end}
      \If{$c^* = C'[i]$} \textbf{continue} \label{alg:leidenlm--best-community-same}
      \EndIf
      \State $\Sigma'[C'[i]] -= K'[i]$ \textbf{;} $\Sigma'[c^*] += K'[i]$ \textbf{atomically} \label{alg:leidenlm--perform-move-begin}
      \State $C'[i] \gets c^*$ \textbf{;} $\Delta Q \gets \Delta Q + \delta Q^*$ \label{alg:leidenlm--perform-move-end}
      \State Mark neighbors of $i$ as unprocessed \label{alg:leidenlm--remark}
    \EndFor \label{alg:leidenlm--loop-vertices-end}
    \If{$\Delta Q \le \tau$} \textbf{break} \Comment{Locally converged?} \label{alg:leidenlm--locally-converged}
    \EndIf
  \EndFor \label{alg:leidenlm--iterations-end}
  \Return{$l_i$} \label{alg:leidenlm--return}
\EndFunction \label{alg:leidenlm--move-end}

\Statex

\Function{scanCommunities}{$H_t, G', C', i, self$}
  \ForAll{$(j, w) \in G'.edges(i)$}
    \If{\textbf{not} $self$ \textbf{and} $i = j$} \textbf{continue}
    \EndIf
    \State $H_t[C'[j]] \gets H_t[C'[j]] + w$
  \EndFor
  \Return{$H_t$}
\EndFunction
\end{algorithmic}
\end{algorithm}

\begin{algorithm}[hbtp]
\caption{Refinement phase of GVE-Leiden.}
\label{alg:leidenre}
\begin{algorithmic}[1]
\Require{$G'$: Input/super-vertex graph}
\Require{$C'$: Community membership of each vertex}
\Require{$K'$: Total edge weight of each vertex}
\Require{$\Sigma'$: Total edge weight of each community}
\Ensure{$H_t$: Collision-free per-thread hashtable}

\Statex

\Function{leidenRefine}{$G', C'_B, C', K', \Sigma'$} \label{alg:leidenre--move-begin}
  \ForAll{$i \in V'$ \textbf{in parallel}} \label{alg:leidenre--loop-vertices-begin}
    \State $c \gets C'[i]$
    \If{$\Sigma'[c] \neq K'[i]$} \textbf{continue} \label{alg:leidenre--check-isolated}
    \EndIf
    \State $H_t \gets scanBounded(\{\}, G', C'_B, C', i, false)$ \label{alg:leidenre--scan}
    \State $\rhd$ Use $H_t, K', \Sigma'$ to choose best community
    \State $c^* \gets$ Best community linked to $i$ in $G'$ within $C'_B$ \label{alg:leidenre--best-community-begin}
    \State $\delta Q^* \gets$ Delta-modularity of moving $i$ to $c^*$ \label{alg:leidenre--best-community-end}
    \If{$c^* = c$} \textbf{continue} \label{alg:leidenre--best-community-same}
    \EndIf
    \If{$atomicCAS(\Sigma'[c], K'[i], 0) = K'[i]$} \label{alg:leidenre--perform-move-begin}
      \State $\Sigma'[c^*] += K'[i]$ \textbf{atomically} \textbf{;} $C'[i] \gets c^*$ \label{alg:leidenre--perform-move-end}
    \EndIf
  \EndFor \label{alg:leidenre--loop-vertices-end}
\EndFunction \label{alg:leidenre--move-end}

\Statex

\Function{scanBounded}{$H_t, G', C'_B, C', i, self$}
  \ForAll{$(j, w) \in G'.edges(i)$}
    \If{\textbf{not} $self$ \textbf{and} $i = j$} \textbf{continue}
    \EndIf
    \If{$C'_B[i] \neq C'_B[j]$} \textbf{continue}
    \EndIf
    \State $H_t[C'[j]] \gets H_t[C'[j]] + w$
  \EndFor
  \Return{$H_t$}
\EndFunction

\Statex

\Function{atomicCAS}{$pointer, old, new$}
  \State $\rhd$ Perform the following atomically
  \If{$pointer = old$} $pointer \gets new$ \textbf{;} \ReturnInline{$old$}
  \Else\ \ReturnInline{$pointer$}
  \EndIf
\EndFunction
\end{algorithmic}
\end{algorithm}

\begin{algorithm}[hbtp]
\caption{Aggregation phase of GVE-Leiden.}
\label{alg:leidenag}
\begin{algorithmic}[1]
\Require{$G'$: Input/super-vertex graph}
\Require{$C'$: Community membership of each vertex}
\Ensure{$G'_{C'}$: Community vertices (CSR)}
\Ensure{$G''$: Super-vertex graph (weighted CSR)}
\Ensure{$*.offsets$: Offsets array of a CSR graph}
\Ensure{$H_t$: Collision-free per-thread hashtable}

\Statex

\Function{leidenAggregate}{$G', C'$}
  \State $\rhd$ Obtain vertices belonging to each community
  \State $G'_{C'}.offsets \gets countCommunityVertices(G', C')$ \label{alg:leidenag--coff-begin}
  \State $G'_{C'}.offsets \gets exclusiveScan(G'_{C'}.offsets)$ \label{alg:leidenag--coff-end}
  \ForAll{$i \in V'$ \textbf{in parallel}} \label{alg:leidenag--comv-begin}
    \State Add edge $(C'[i], i)$ to CSR $G'_{C'}$ \textbf{atomically}
  \EndFor \label{alg:leidenag--comv-end}
  \State $\rhd$ Obtain super-vertex graph
  \State $G''.offsets \gets communityTotalDegree(G', C')$ \label{alg:leidenag--yoff-begin}
  \State $G''.offsets \gets exclusiveScan(G''.offsets)$ \label{alg:leidenag--yoff-end}
  \State $|\Gamma| \gets$ Number of communities in $C'$
  \ForAll{$c \in [0, |\Gamma|)$ \textbf{in parallel}} \label{alg:leidenag--y-begin}
    \State $H_t \gets \{\}$
    \ForAll{$i \in G'_{C'}.edges(c)$}
      \State $H_t \gets scanCommunities(H_t, G', C', i, true)$
    \EndFor
    \ForAll{$(d, w) \in H_t$}
      \State Add edge $(c, d, w)$ to CSR $G''$ \textbf{atomically}
    \EndFor
  \EndFor \label{alg:leidenag--y-end}
  \Return $G''$ \label{alg:leidenag--return}
\EndFunction
\end{algorithmic}
\end{algorithm}

\subsubsection{Local-moving phase of GVE-Leiden}

The pseuodocode for the local-moving phase of GVE-Leiden is shown in Algorithm \ref{alg:leidenlm}, which iteratively moves vertices between communities to maximize modularity. Here, the \texttt{leidenMove()} function takes the current graph $G'$, community membership $C'$, total edge weight of each vertex $K'$ and each community $\Sigma'$, the iteration tolerance $\tau$ as input, and returns the number of iterations performed $l_i$.

Lines \ref{alg:leidenlm--iterations-begin}-\ref{alg:leidenlm--iterations-end} represent the main loop of the local-moving phase. In line \ref{alg:leidenlm--reset-affected}, we first mark all vertices as unprocessed. Then, in line \ref{alg:leidenlm--init-deltaq}, we initialize the total delta-modularity per iteration $\Delta Q$. Next, in lines \ref{alg:leidenlm--loop-vertices-begin}-\ref{alg:leidenlm--loop-vertices-end}, we iterate over unprocessed vertices in parallel. For each unprocessed vertex $i$, we mark $i$ as processed - vertex pruning (line \ref{alg:leidenlm--prune}), scan communities connected to $i$ - excluding self (line \ref{alg:leidenlm--scan}), determine the best community $c*$ to move $i$ to (line \ref{alg:leidenlm--best-community-begin}), and calculate the delta-modularity of moving $i$ to $c*$ (line \ref{alg:leidenlm--best-community-end}). We then update the community membership of $i$ (lines \ref{alg:leidenlm--perform-move-begin}-\ref{alg:leidenlm--perform-move-end}) and mark its neighbors as unprocessed (line \ref{alg:leidenlm--remark}) if a better community was found. In line \ref{alg:leidenlm--locally-converged}, we check if the local-moving phase has converged. If so, we break out of the loop (or if $MAX\_ITERATIONS$ is reached). At the end, in line \ref{alg:leidenlm--return}, we return the number of iterations performed $l_i$.

\subsubsection{Refinement phase of GVE-Leiden}

The pseuodocode for the refinement phase of GVE-Leiden is presented in Algorithm \ref{alg:leidenre}. This is similar to the local-moving phase, but utilizes the obtained community membership of each vertex as a \textit{community bound}, where each vertex must choose to join the community of another vertex within its community bound.\ignore{Similar to the local-moving phase however, vertices iteratively move between communities to maximize modularity.} At the start of the refinement phase, the community membership of each vertex is reset, such that each vertex belongs to its own community. Here, the \texttt{leidenRefine()} function takes the current graph $G'$, the community bound of each vertex $C'_B$, the initial community membership $C'$ of each vertex, the total edge weight of each vertex $K'$, the initial total edge weight of each community $\Sigma'$, and the current per iteration tolerance $\tau$ as input, and returns the number of iterations performed $l_j$.

Lines \ref{alg:leidenre--loop-vertices-begin}-\ref{alg:leidenre--loop-vertices-end} represent the core of the refinement phase. In the refinement phase, we perform, what is called the constrained merge procedure \cite{com-traag19}. The idea here is to allow vertices, within each community bound, to form sub-communities by only allowing isolated vertices (i.e., vertices belonging to their own community) to change their community membership. This procedure splits any internally-disconnected communities identified during the local-moving phase, and prevents the formation of any new disconnected communities. Here, for each isolated vertex $i$ (line \ref{alg:leidenre--check-isolated}), we scan communities connected to $i$ within the \textit{same community bound} - excluding self (line \ref{alg:leidenre--scan}), evaluate the best community $c*$ to move $i$ to (line \ref{alg:leidenre--best-community-begin}), and compute the delta-modularity of moving $i$ to $c*$ (line \ref{alg:leidenre--best-community-end}). If a better community was found, we attempt to update the community membership of $i$ if it is still isolated (lines \ref{alg:leidenre--perform-move-begin}-\ref{alg:leidenre--perform-move-end}).

\subsubsection{Aggregation phase of GVE-Leiden}

Finally, we show the psuedocode for the aggregation phase in Algorithm \ref{alg:leidenag}, where communities are aggregated into super-vertices in preparation for the next pass of the Leiden algorithm (which operates on the super-vertex graph). Here, the \texttt{leidenAggregate()} function takes the current graph $G'$ and the community membership $C'$ as input, and returns the super-vertex graph $G''$.

In lines \ref{alg:leidenag--coff-begin}-\ref{alg:leidenag--coff-end}, the offsets array for the community vertices CSR $G'_{C'}.offsets$ is obtained. This is achieved by initially counting the number of vertices in each community using \texttt{countCommunityVert} \texttt{ices()} and subsequently performing an exclusive scan on the array. In lines \ref{alg:leidenag--comv-begin}-\ref{alg:leidenag--comv-end},  a parallel iteration over all vertices is performed to atomically populate vertices belonging to each community into the community graph CSR $G'_{C'}$. Following this, the offsets array for the super-vertex graph CSR is obtained by overestimating the degree of each super-vertex. This involves calculating the total degree of each community with \texttt{communityTotalDegree()} and performing an exclusive scan on the array (lines \ref{alg:leidenag--yoff-begin}-\ref{alg:leidenag--yoff-end}). As a result, the super-vertex graph CSR becomes holey, featuring gaps between the edges and weights arrays of each super-vertex in the CSR.

Then, in lines \ref{alg:leidenag--y-begin}-\ref{alg:leidenag--y-end}, a parallel iteration over all communities $c \in [0, |\Gamma|)$ is performed. For each vertex $i$ belonging to community $c$, all communities $d$ (with associated edge weight $w$), linked to $i$ as defined by \texttt{scanCommunities()} in Algorithm \ref{alg:leidenlm}, are added to the per-thread hashtable $H_t$. Once $H_t$ is populated with all communities (and associated weights) linked to community $c$, these are atomically added as edges to super-vertex $c$ in the super-vertex graph $G''$. Finally, in line \ref{alg:leidenag--return}, we return the super-vertex graph $G''$.

\section{Evaluation}
\label{sec:evaluation}
\subsection{Experimental Setup}
\label{sec:setup}

\subsubsection{System used}

We employ a server equipped with two Intel Xeon Gold 6226R processors, each featuring $16$ cores running at a clock speed of $2.90$ GHz. Each core is equipped with a $1$ MB L1 cache, a $16$ MB L2 cache, and a $22$ MB shared L3 cache. The system is configured with $376$ GB RAM and set up with CentOS Stream 8. For our GPU experiments, we utilize a system with an NVIDIA A100 GPU (108 SMs, 64 CUDA cores per SM, 80 GB global memory, 1935 GB/s bandwidth, 164 KB shared memory per SM) and an AMD EPYC-7742 processor (64 cores, 2.25 GHz). The server is equipped with 512 GB DDR4 RAM and runs Ubuntu 20.04.

\subsubsection{Configuration}

We use 32-bit integers for vertex ids and 32-bit float for edge weights but use 64-bit floats for computations and hashtable values. We utilize $64$ threads to match the number of cores available on the system (unless specified otherwise). For compilation, we use GCC 8.5 and OpenMP 4.5 on the CPU system, and GCC 9.4, OpenMP 5.0, and CUDA 11.4 on the GPU system.

\subsubsection{Dataset}

The graphs used in our experiments are given in Table \ref{tab:dataset}. These are sourced from the SuiteSparse Matrix Collection \cite{suite19}. In the graphs, number of vertices vary from $3.07$ to $214$ million, and number of edges vary from $25.4$ million to $3.80$ billion. We ensure edges to be undirected and weighted with a default of $1$.

\begin{table}[hbtp]
  \centering
  \caption{List of $13$ graphs obtained SuiteSparse Matrix Collection \cite{suite19} (directed graphs are marked with $*$). Here, $|V|$ is the number of vertices, $|E|$ is the number of edges (after adding reverse edges), $D_{avg}$ is the average degree, and $|\Gamma|$ is the number of communities obtained with \textit{GVE-Leiden}.}
  \label{tab:dataset}
  \begin{tabular}{|c||c|c|c|c|}
    \toprule
    \textbf{Graph} &
    \textbf{\textbf{$|V|$}} &
    \textbf{\textbf{$|E|$}} &
    \textbf{\textbf{$D_{avg}$}} &
    \textbf{\textbf{$|\Gamma|$}} \\
    \midrule
    \multicolumn{5}{|c|}{\textbf{Web Graphs (LAW)}} \\ \hline
    indochina-2004$^*$ & 7.41M & 341M & 41.0 & 2.68K \\ \hline
    uk-2002$^*$ & 18.5M & 567M & 16.1 & 41.8K \\ \hline
    arabic-2005$^*$ & 22.7M & 1.21B & 28.2 & 2.92K \\ \hline
    uk-2005$^*$ & 39.5M & 1.73B & 23.7 & 18.2K \\ \hline
    webbase-2001$^*$ & 118M & 1.89B & 8.6 & 2.94M \\ \hline
    it-2004$^*$ & 41.3M & 2.19B & 27.9 & 4.05K \\ \hline
    sk-2005$^*$ & 50.6M & 3.80B & 38.5 & 2.67K \\ \hline
    \multicolumn{5}{|c|}{\textbf{Social Networks (SNAP)}} \\ \hline
    com-LiveJournal & 4.00M & 69.4M & 17.4 & 3.09K \\ \hline
    com-Orkut & 3.07M & 234M & 76.2 & 36 \\ \hline
    \multicolumn{5}{|c|}{\textbf{Road Networks (DIMACS10)}} \\ \hline
    asia\_osm & 12.0M & 25.4M & 2.1 & 2.70K \\ \hline
    europe\_osm & 50.9M & 108M & 2.1 & 6.13K \\ \hline
    \multicolumn{5}{|c|}{\textbf{Protein k-mer Graphs (GenBank)}} \\ \hline
    kmer\_A2a & 171M & 361M & 2.1 & 21.1K \\ \hline
    kmer\_V1r & 214M & 465M & 2.2 & 10.5K \\ \hline
  \bottomrule
  \end{tabular}
\end{table}

\begin{figure*}[hbtp]
  \centering
  \subfigure[Runtime in seconds (logarithmic scale) with \textit{Original Leiden}, \textit{igraph Leiden}, \textit{NetworKit Leiden}, \textit{cuGraph Leiden}, and \textit{GVE-Leiden}]{
    \label{fig:leiden-compare--runtime}
    \includegraphics[width=0.98\linewidth]{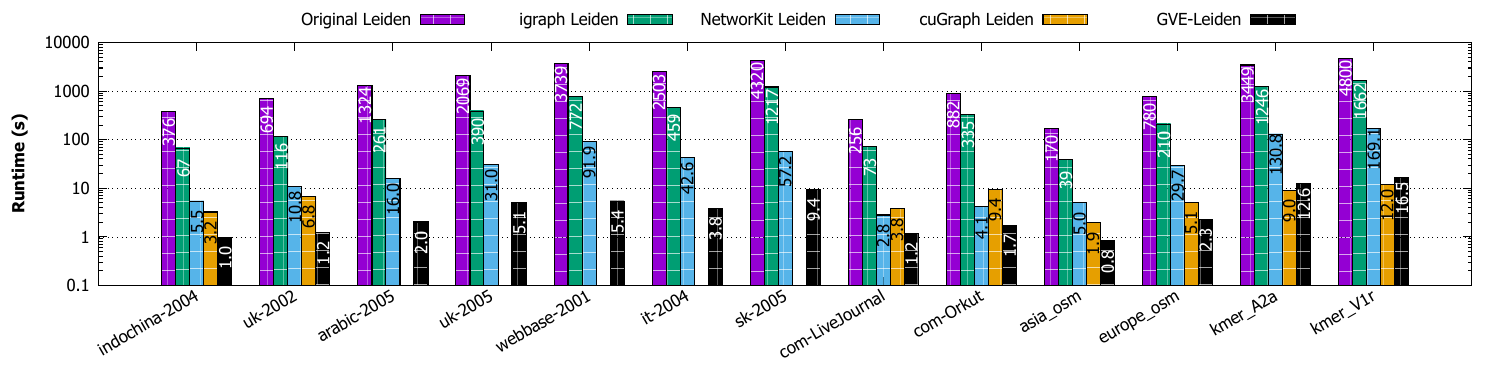}
  } \\[-0ex]
  \subfigure[Speedup of \textit{GVE-Leiden} (logarithmic scale) with respect to \textit{Original Leiden}, \textit{igraph Leiden}, \textit{NetworKit Leiden}, and \textit{cuGraph Leiden}.]{
    \label{fig:leiden-compare--speedup}
    \includegraphics[width=0.98\linewidth]{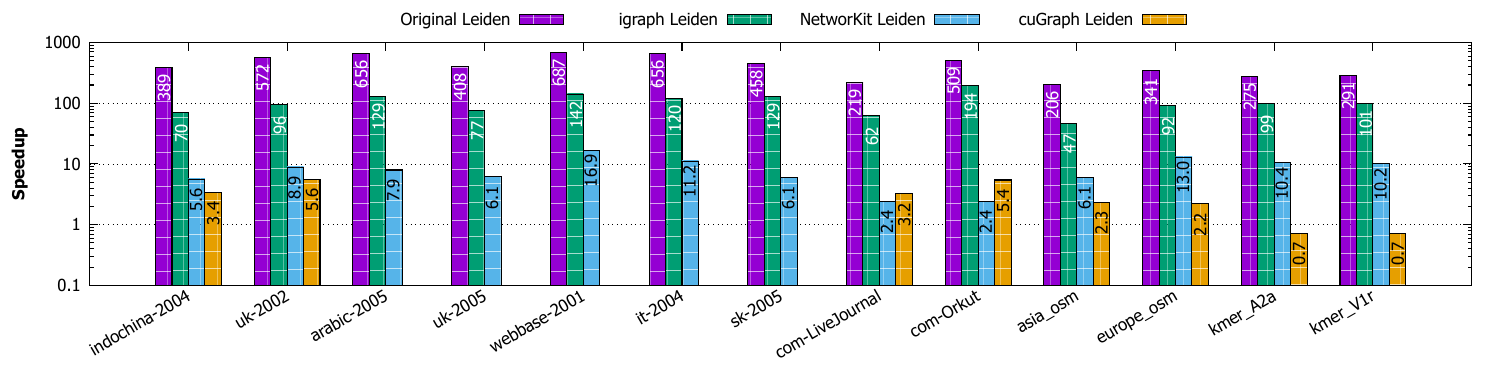}
  } \\[-0ex]
  \subfigure[Modularity of communities obtained with \textit{Original Leiden}, \textit{igraph Leiden}, \textit{NetworKit Leiden}, \textit{cuGraph Leiden}, and \textit{GVE-Leiden}.]{
    \label{fig:leiden-compare--modularity}
    \includegraphics[width=0.98\linewidth]{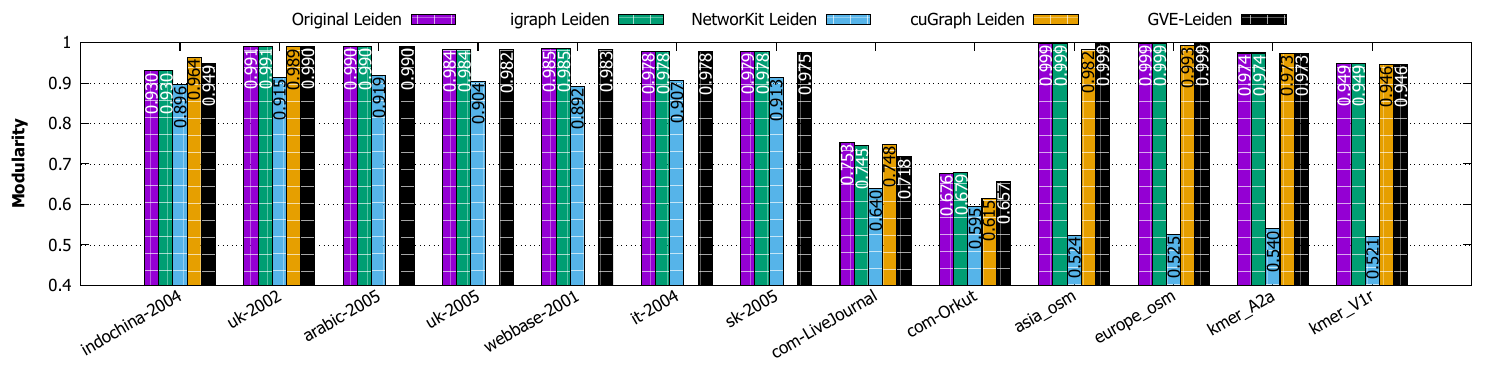}
  } \\[-0ex]
  \subfigure[Fraction of disconnected communities (logarithmic scale) with \textit{Original Leiden}, \textit{igraph Leiden}, \textit{NetworKit Leiden}, \textit{cuGraph Leiden}, and \textit{GVE-Leiden}.]{
    \label{fig:leiden-compare--disconnected}
    \includegraphics[width=0.98\linewidth]{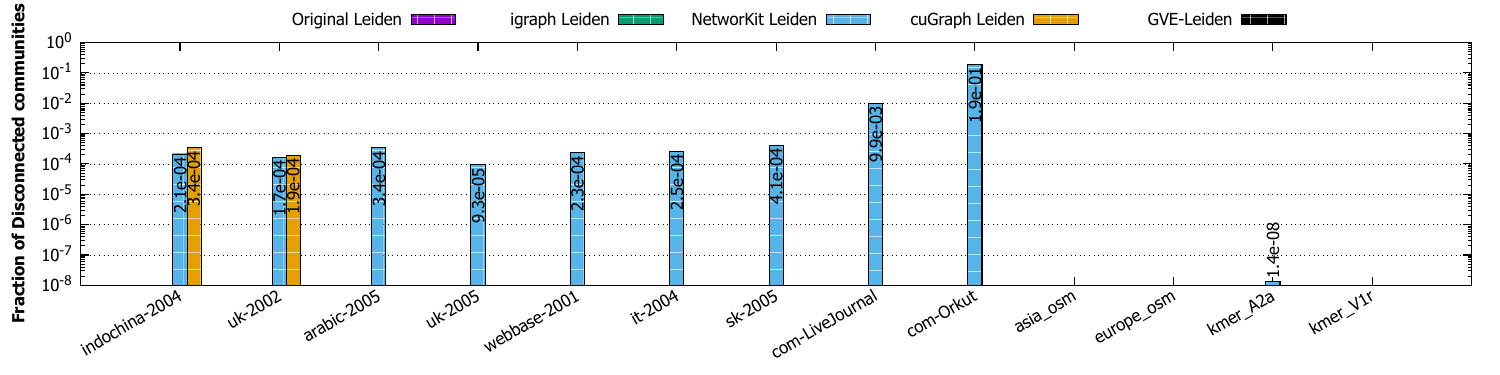}
  } \\[-2ex]
  \caption{Runtime in seconds (log-scale), speedup (log-scale), modularity, and fraction of disconnected communities (log-scale) with \textit{Original Leiden}, \textit{igraph Leiden}, \textit{NetworKit Leiden}, \textit{cuGraph Leiden}, and \textit{GVE-Leiden} for each graph in the dataset.}
  \label{fig:leiden-compare}
\end{figure*}

\begin{figure*}[hbtp]
  \centering
  \subfigure[Runtime in seconds (logarithmic scale) with \textit{GVE-Louvain} and \textit{GVE-Leiden}]{
    \label{fig:gve-compare--runtime}
    \includegraphics[width=0.98\linewidth]{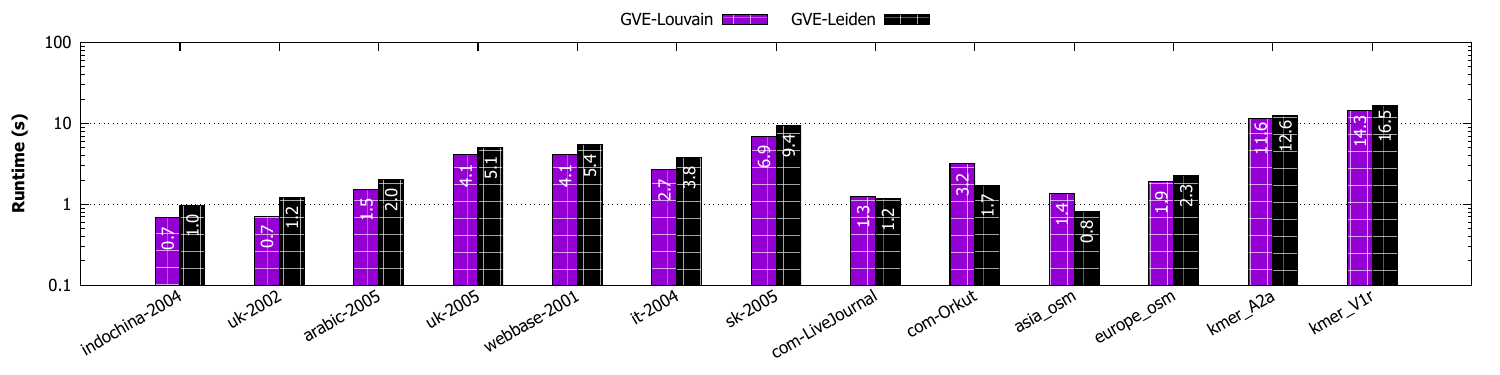}
  } \\[-0ex]
  \subfigure[Speedup of \textit{GVE-Leiden} with respect to \textit{GVE-Louvain}. \textit{GVE-Leiden} is generally slower (speedup < $1$) because of additional refinement phase.]{
    \label{fig:gve-compare--speedup}
    \includegraphics[width=0.98\linewidth]{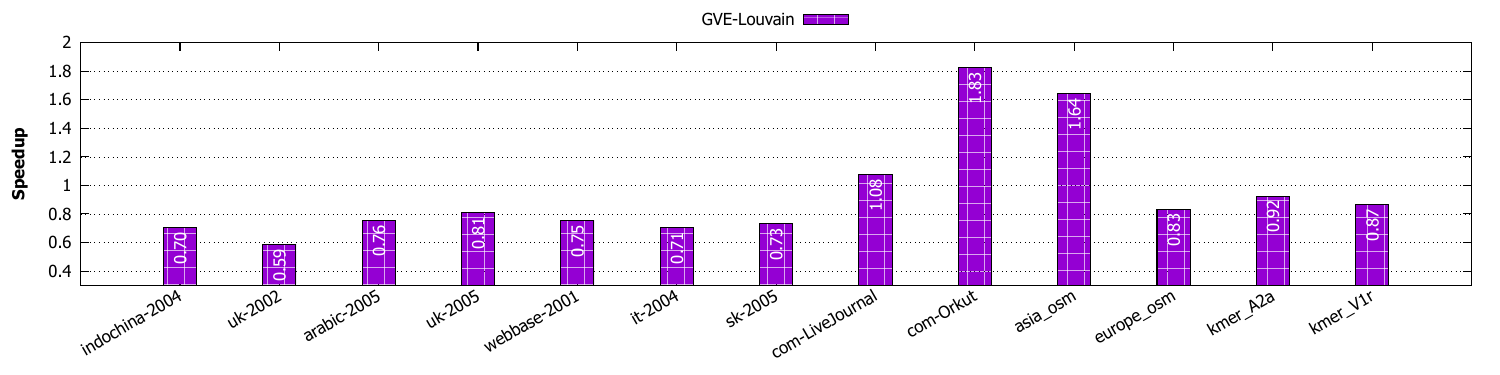}
  } \\[-0ex]
  \subfigure[Modularity of communities obtained with \textit{GVE-Louvain} and \textit{GVE-Leiden}.]{
    \label{fig:gve-compare--modularity}
    \includegraphics[width=0.98\linewidth]{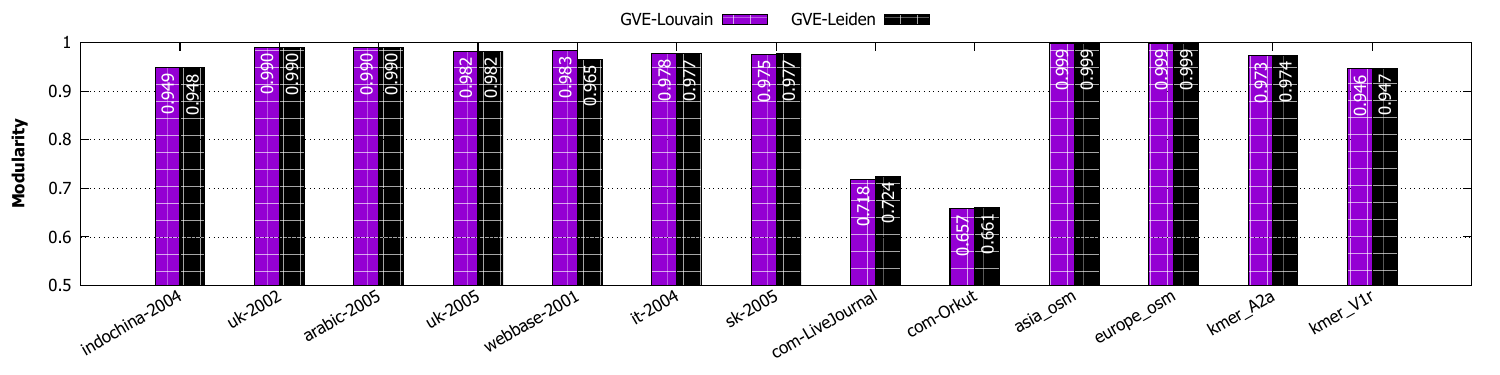}
  } \\[-0ex]
  \subfigure[Fraction of disconnected communities (logarithmic scale) with \textit{GVE-Louvain} and \textit{GVE-Leiden}.]{
    \label{fig:gve-compare--disconnected}
    \includegraphics[width=0.98\linewidth]{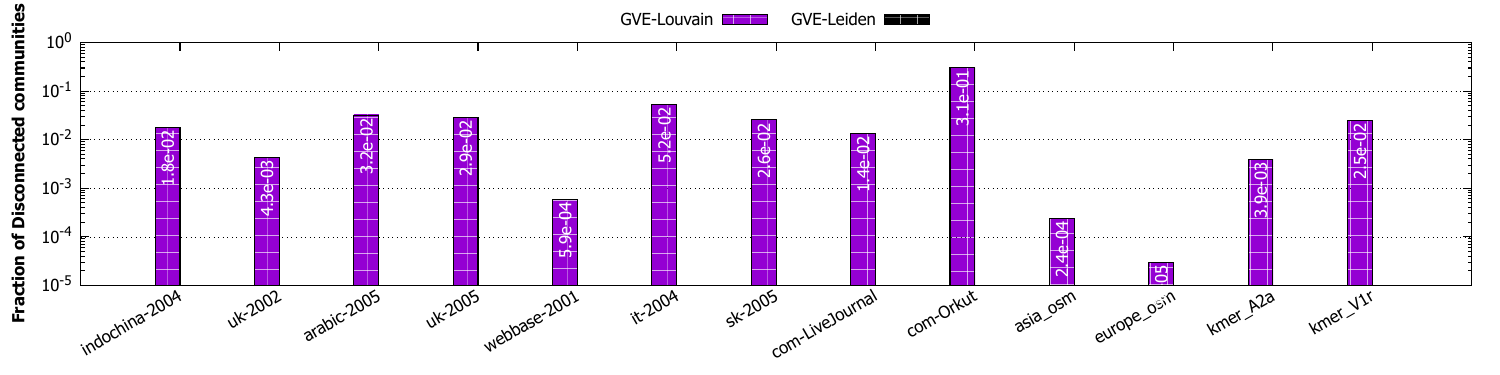}
  } \\[-2ex]
  \caption{Runtime in seconds (log-scale), speedup, modularity, and fraction of disconnected communities (log-scale) with \textit{GVE-Louvain} and \textit{GVE-Leiden} for each graph in the dataset.}
  \label{fig:gve-compare}
\end{figure*}

\subsection{Comparing Performance of GVE-Leiden}
\label{sec:comparison}

We now compare the performance of GVE-Leiden with the original Leiden \cite{com-traag19}, igraph Leiden \cite{csardi2006igraph}, NetworKit Leiden \cite{staudt2016networkit}, and cuGraph Leiden \cite{kang2023cugraph}. The original Leiden and igraph Leiden are sequential implementations, while NetworKit Leiden and GVE-Leiden are parallel multicore implementations of the Leiden algorithm.\ignore{These are run on our CPU-only system.} On the other hand, cuGraph Leiden is a GPU implementation of Leiden, and is run on our GPU-based system. For the original Leiden, we use a C++ program to initialize a \texttt{ModularityVertexPartition} upon the loaded graph, and invoke \texttt{optimise\_partition()} to obtain the community membership of each vertex in the graph. On graphs with a large number of edges, such as \textit{webbase-2001} and \textit{sk-2005}, using \texttt{ModularityVertexPartition} introduces disconnected communities due to issues with numerical precision (i.e. the improvement of separating two disconnected parts may be positive, but due to the enormous weight, this may effectively be near 0) \cite{traag2024leiden}. For such graphs, we instead use \texttt{RBConfigurationVertexPartition}, which uses unscaled improvements to modularity (i.e. they do not scale with the total weight)\ignore{ --- the problem of disconnected communities is not observed here}. For igraph Leiden, we use \texttt{igraph\_comm} \texttt{unity\_leiden()} with a resolution of $1/2|E|$, a beta of $0.01$, and request the algorithm to run until convergence. For NetworKit Leiden, we write a Python script to call \texttt{ParallelLeiden()}, while limiting the number of passes to $10$. For cuGraph Leiden, we write a Python script to configure the Rapids Memory Manager (RMM) to use a pool allocator for fast memory allocations and run \texttt{cugraph.leiden()} on the loaded graph. For each graph, we measure the runtime of each implementation and the modularity of the communities obtained, five times, for averaging. With cuGraph, we disregard the runtime of the first run to ensure subsequent measurements only reflect RMM's pool usage without CUDA memory allocation overhead. We save the obtained community membership vector to a file and later count the disconnected communities using an algorithm given in our extended report \cite{report}. In all instances, we use modularity as the quality function to optimize for.

Figure \ref{fig:leiden-compare--runtime} shows the runtimes of the original Leiden, igraph Leiden, NetworKit Leiden, cuGraph Leiden, and GVE-Leiden on each graph in the dataset. cuGraph Leiden fails to run on the \textit{arabic-2005}, \textit{uk-2005}, \textit{webbase-2001}, \textit{it-2004}, and \textit{sk-2005} graphs due to out of memory issues. On the \textit{sk-2005} graph, GVE-Leiden finds communities in $9.4$ seconds, and thus achieve a processing rate of $403$ million edges/s. Figure \ref{fig:leiden-compare--speedup} shows the speedup of GVE-Leiden with respect to each implementation mentioned above. GVE-Leiden is on average $436\times$, $104\times$, $8.2\times$, and $3.0\times$ faster than the original Leiden, igraph Leiden, NetworKit Leiden, and cuGraph Leiden respectively. Figure \ref{fig:leiden-compare--modularity} shows the modularity of communities obtained with each implementation. GVE-Leiden on average obtains $0.3\%$ lower modularity than the original Leiden and igraph Leiden, $25\%$ higher modularity than NetworKit Leiden (especially on road networks and protein k-mer graphs), and $3.5\%$ higher modularity that cuGraph Leiden (primarily due to cuGraph Leiden's inability to run on well-clusterable graphs). Finally, Figure \ref{fig:leiden-compare--disconnected} shows the fraction of disconnected communities obtained with each implementation. Here, the absence of bars indicates the absence of disconnected communities. Communities identified by NetworKit Leiden and cuGraph Leiden have on average $1.5\times10^{-2}$ and $6.6\times10^{-5}$ fraction of disconnected communities, respectively, while none of the communities identified by the original Leiden, igraph Leiden, and GVE-Leiden are internally-disconnected. As the Leiden algorithm guarantees the absence of disconnected communities \cite{com-traag19}, those observed with NetworKit Leiden and cuGraph Leiden are likely due to implementation issues.

\ignore{
Q> Why does Leiden have some disconnected communities still?
Q> Why does Leiden not work with road networks?
I think the answer for both questions has to do with parallelism. There is a race between threads to pick a suitable community. If the size of each community is large, this generally does not badly affect modularity. But with small community bounds (in refinement phase of Leiden) the race can lead to bad community memberships. I will try to come up a few solutions to this. This is not an issue with sequential, so Traag et al. dont observe this issue in their original paper.
A> This is now resolved!
}

\begin{figure*}[hbtp]
  \centering
  \subfigure[Phase split]{
    \label{fig:leiden-splits--phase}
    \includegraphics[width=0.48\linewidth]{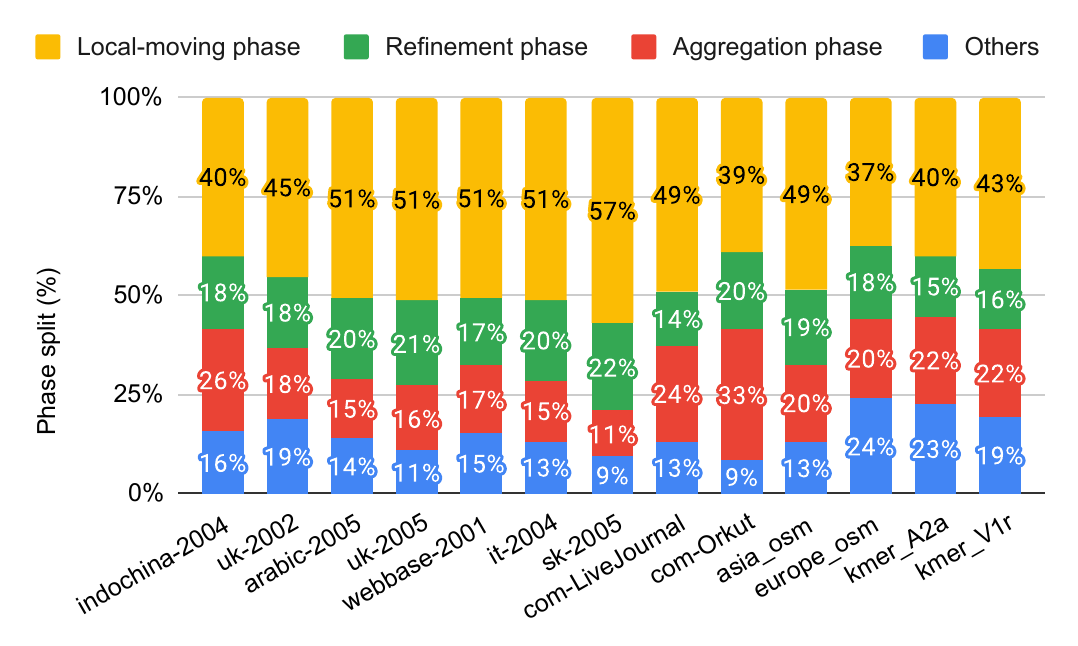}
  }
  \subfigure[Pass split]{
    \label{fig:leiden-splits--pass}
    \includegraphics[width=0.48\linewidth]{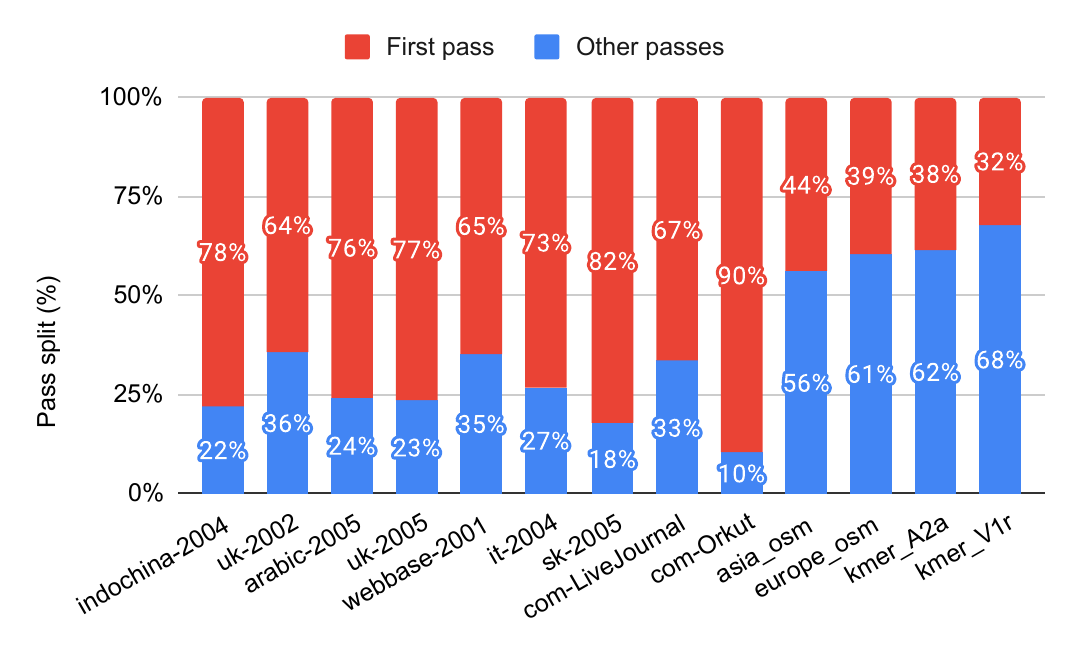}
  } \\[-2ex]
  \caption{Phase split of \textit{GVE-Leiden} shown on the left, and pass split shown on the right for each graph in the dataset.}
  \label{fig:leiden-splits}
\end{figure*}

\begin{figure}[hbtp]
  \centering
  \subfigure{
    \label{fig:leiden-hardness--all}
    \includegraphics[width=0.98\linewidth]{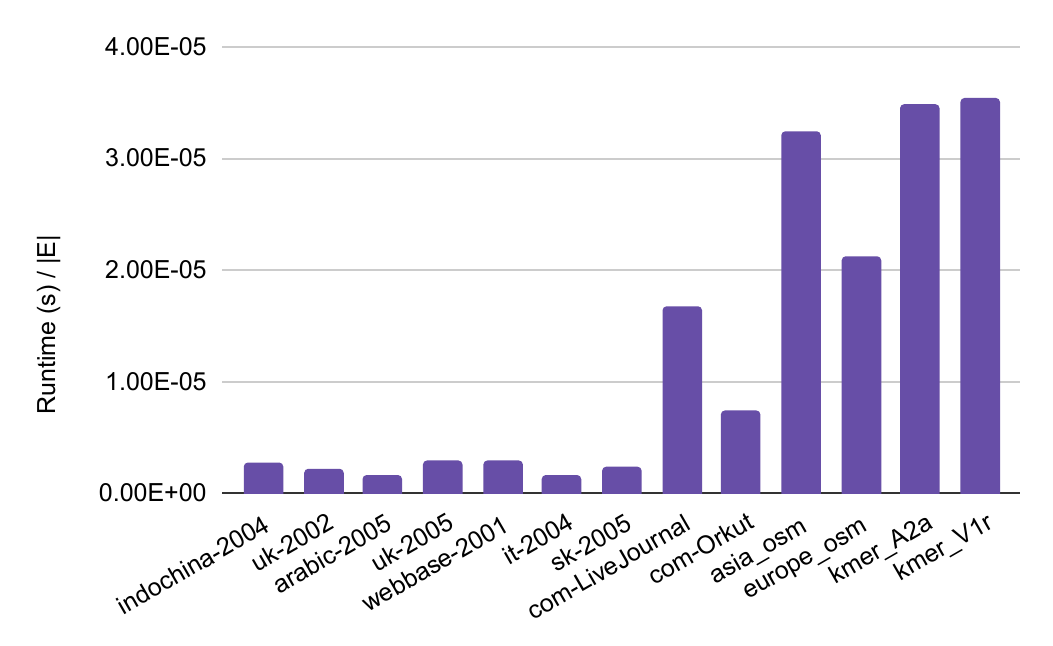}
  } \\[-2ex]
  \caption{Runtime $/ |E|$ factor with \textit{GVE-Leiden} for each graph\ignore{in the dataset}.}
  \label{fig:leiden-hardness}
\end{figure}

\begin{figure}[hbtp]
  \centering
  \includegraphics[width=0.98\linewidth]{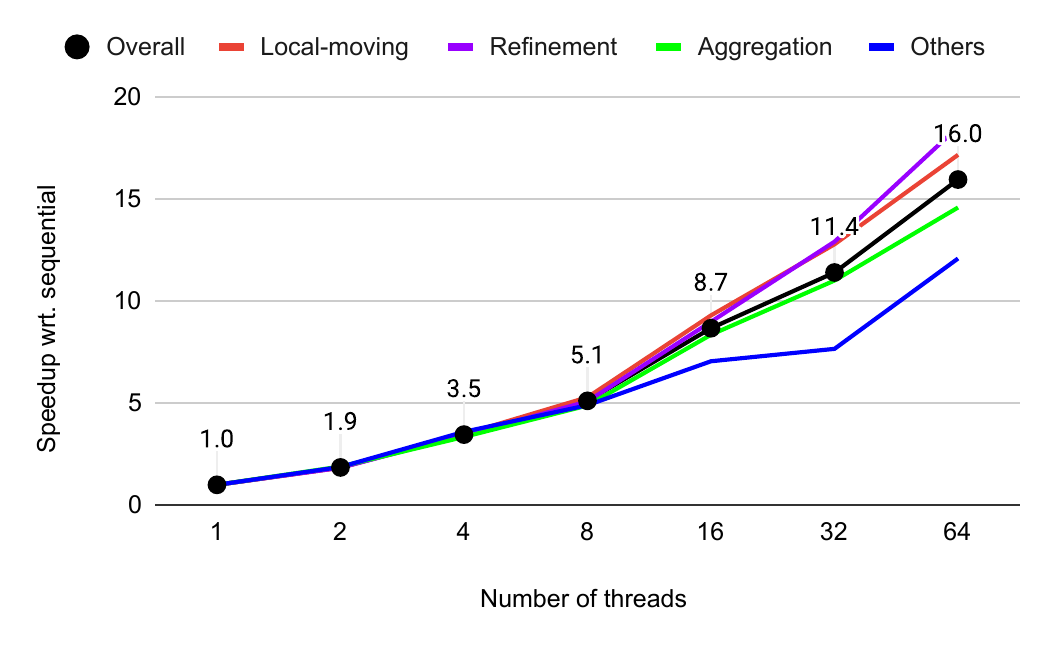} \\[-2ex]
  \caption{Overall speedup of \textit{GVE-Leiden}, and its various phases (local-moving, refinement, aggregation, others), with increasing number of threads (in multiples of 2).}
  \label{fig:leiden-ss}
\end{figure}

Next, we compare the performance of GVE-Leiden with GVE-Louvain \cite{sahu2023gvelouvain}, our parallel implementation of the Louvain method. As above, for each graph in the dataset, we run both algorithms 5 times to minimize measurement noise\ignore{, and report the averages in Figures \ref{fig:gve-compare--runtime}, \ref{fig:gve-compare--speedup}, \ref{fig:gve-compare--modularity}, and \ref{fig:gve-compare--disconnected}}. Figure \ref{fig:gve-compare--runtime} shows the runtimes of GVE-Louvain and GVE-Leiden on each graph in the dataset. Figure \ref{fig:gve-compare--speedup} shows the speedup of GVE-Leiden with respect to GVE-Louvain. GVE-Leiden is on average $13\%$ slower than GVE-Louvain. This increase in computation time is a trade-off for identifying communities that are not internally-disconnected, as given below. Figure \ref{fig:gve-compare--modularity} shows the modularity of communities obtained with GVE-Louvain and GVE-Leiden. GVE-Leiden on average obtains the same modularity as GVE-Louvain. Finally, Figure \ref{fig:gve-compare--disconnected} shows the fraction of internally-disconnected communities obtained\ignore{with GVE-Louvain and GVE-Leiden}. Communities identified by GVE-Louvain on average have $4.0\%$ disconnected communities, while GVE-Leiden has none.

\subsection{Analyzing Performance of GVE-Leiden}

We now analyze the performance of GVE-Leiden. The phase-wise and pass-wise split of GVE-Leiden is shown in Figures \ref{fig:leiden-splits--phase} and \ref{fig:leiden-splits--pass} respectively. Figure \ref{fig:leiden-splits--phase} reveals that GVE-Leiden devotes a significant portion of its runtime to the local-moving and refinement phases on \textit{web graphs}, \textit{road networks}, and \textit{protein k-mer graphs}, while it dedicates majority of its runtime in the aggregation phase on \textit{social networks}. The pass-wise split, shown in Figure \ref{fig:leiden-splits--pass}, indicates that the first pass is time-intensive for high-degree graphs (\textit{web graphs} and \textit{social networks}), while subsequent passes take precedence in execution time on low-degree graphs (\textit{road networks} and \textit{protein k-mer graphs}).

On average, GVE-Leiden spends $46\%$ of its runtime in the local-moving phase, $19\%$ in the refinement phase, $20\%$ in the aggregation phase, and $15\%$ in other steps (initialization, renumbering communities, dendrogram lookup, and resetting communities). Further, $63\%$ of the runtime is consumed by the first pass of the algorithm, which is computationally demanding due to the size of the original graph (subsequent passes operate on super-vertex graphs). We also observe that graphs with lower average degree (\textit{road networks} and \textit{protein k-mer graphs}) and those with poor community structure (e.g., \verb|com-LiveJournal| and \verb|com-Orkut|) exhibit a higher $\text{runtime}/|E|$ factor, as shown in Figure \ref{fig:leiden-hardness}.

\subsection{Strong Scaling of GVE-Leiden}

Finally, we assess the strong scaling performance of GVE-Leiden. In this analysis, we vary the number of threads from $1$ to $64$ in multiples of $2$ for each input graph, and measure the total time taken for GVE-Leiden to identify communities, encompassing its phase splits (local-moving, refinement, aggregation, and others), repeated five times for averaging. The results are shown in Figure \ref{fig:leiden-ss}. With 32 threads, GVE-Leiden achieves an average speedup of $11.4\times$ compared to a single-threaded execution, indicating a performance increase of $1.6\times$ for every doubling of threads. Nevertheless, scalability is restricted due to the sequential nature of steps/phases in the algorithm. At 64 threads, GVE-Leiden is affected by NUMA effects, resulting in a speedup of only $16.0\times$.

\section{Conclusion}
\label{sec:conclusion}
In conclusion, this study addresses the design of the most optimized multicore implementation of the Leiden algorithm \cite{com-traag19}, to the best of our knowledge. Here, we extend optimizations from our implementation of the Louvain algorithm \cite{sahu2023gvelouvain}, and use a greedy refinement phase where vertices greedily optimize for delta-modularity within their community bounds, which we observe, offers both better performance and quality than a randomized approach. On a system equipped with two 16-core Intel Xeon Gold 6226R processors, our implementation of the Leiden algorithm, referred to as GVE-Leiden, attains a processing rate of $403 M$ edges per second on a $3.8 B$ edge graph. It outperforms the original Leiden implementation, igraph Leiden, NetworKit Leiden, and cuGraph Leiden (run on an NVIDIA A100 GPU) by $436\times$, $104\times$, $8.2\times$, and $3.0\times$ respectively. GVE-Leiden identifies communities of equivalent quality to the first two implementations, and $25\%$ / $3.5\%$ higher quality than NetworKit / cuGraph. Doubling the number of threads results in a average performance scaling of $1.6\times$ for GVE-Leiden.

In a previous version of this report, we implemented the refinement phase of the Leiden algorithm utilizing a \textit{constrained move} procedure, which does not guarantee the absence of disconnected communities. In this current version of the report, we have transitioned to employing the \textit{constrained merge} procedure alongside atomics to ensure no internally-disconnected communities. We also addressed issues in measuring disconnected communities for the original Leiden and igraph Leiden, which arose due to the number of vertices in a graph varying between the Matrix Market and the Edgelist formats (which does not have isolated vertices), and used the \texttt{RBConfigurationVertexPartition} with the original Leiden for large graphs (i.e., \textit{webbase-2001} and \textit{sk-2005}).

\begin{acks}
I would like to thank Prof. Kishore Kothapalli, Prof. Dip Sankar Banerjee, Vincent Traag, Geerten Verweij, Fabian Nguyen, Chuck Hastings, and Rick Ratzel for their support.
\end{acks}

\bibliographystyle{ACM-Reference-Format}
\bibliography{main}


\begin{thebibliography}{37}


\ifx \showCODEN    \undefined \def \showCODEN     #1{\unskip}     \fi
\ifx \showDOI      \undefined \def \showDOI       #1{#1}\fi
\ifx \showISBNx    \undefined \def \showISBNx     #1{\unskip}     \fi
\ifx \showISBNxiii \undefined \def \showISBNxiii  #1{\unskip}     \fi
\ifx \showISSN     \undefined \def \showISSN      #1{\unskip}     \fi
\ifx \showLCCN     \undefined \def \showLCCN      #1{\unskip}     \fi
\ifx \shownote     \undefined \def \shownote      #1{#1}          \fi
\ifx \showarticletitle \undefined \def \showarticletitle #1{#1}   \fi
\ifx \showURL      \undefined \def \showURL       {\relax}        \fi
\providecommand\bibfield[2]{#2}
\providecommand\bibinfo[2]{#2}
\providecommand\natexlab[1]{#1}
\providecommand\showeprint[2][]{arXiv:#2}

\bibitem[Aldabobi et~al\mbox{.}(2022)]%
        {com-aldabobi22}
\bibfield{author}{\bibinfo{person}{A. Aldabobi}, \bibinfo{person}{A. Sharieh}, {and} \bibinfo{person}{R. Jabri}.} \bibinfo{year}{2022}\natexlab{}.
\newblock \showarticletitle{An improved Louvain algorithm based on Node importance for Community detection}.
\newblock \bibinfo{journal}{\emph{Journal of Theoretical and Applied Information Technology}} \bibinfo{volume}{100}, \bibinfo{number}{23} (\bibinfo{year}{2022}), \bibinfo{pages}{1--14}.
\newblock


\bibitem[Blondel et~al\mbox{.}(2008)]%
        {com-blondel08}
\bibfield{author}{\bibinfo{person}{V. Blondel}, \bibinfo{person}{J. Guillaume}, \bibinfo{person}{R. Lambiotte}, {and} \bibinfo{person}{E. Lefebvre}.} \bibinfo{year}{2008}\natexlab{}.
\newblock \showarticletitle{{Fast unfolding of communities in large networks}}.
\newblock \bibinfo{journal}{\emph{Journal of Statistical Mechanics: Theory and Experiment}} \bibinfo{volume}{2008}, \bibinfo{number}{10} (\bibinfo{date}{Oct} \bibinfo{year}{2008}), \bibinfo{pages}{P10008}.
\newblock


\bibitem[Brandes et~al\mbox{.}(2007)]%
        {com-brandes07}
\bibfield{author}{\bibinfo{person}{U. Brandes}, \bibinfo{person}{D. Delling}, \bibinfo{person}{M. Gaertler}, \bibinfo{person}{R. Gorke}, \bibinfo{person}{M. Hoefer}, \bibinfo{person}{Z. Nikoloski}, {and} \bibinfo{person}{D. Wagner}.} \bibinfo{year}{2007}\natexlab{}.
\newblock \showarticletitle{{On modularity clustering}}.
\newblock \bibinfo{journal}{\emph{IEEE transactions on knowledge and data engineering}} \bibinfo{volume}{20}, \bibinfo{number}{2} (\bibinfo{year}{2007}), \bibinfo{pages}{172--188}.
\newblock


\bibitem[Cheong et~al\mbox{.}(2013)]%
        {com-cheong13}
\bibfield{author}{\bibinfo{person}{C. Cheong}, \bibinfo{person}{H. Huynh}, \bibinfo{person}{D. Lo}, {and} \bibinfo{person}{R. Goh}.} \bibinfo{year}{2013}\natexlab{}.
\newblock \showarticletitle{{Hierarchical Parallel Algorithm for Modularity-Based Community Detection Using GPUs}}. In \bibinfo{booktitle}{\emph{Proceedings of the 19th International Conference on Parallel Processing}} (Aachen, Germany) \emph{(\bibinfo{series}{Euro-Par'13})}. \bibinfo{publisher}{Springer-Verlag}, \bibinfo{address}{Berlin, Heidelberg}, \bibinfo{pages}{775--787}.
\newblock
\showISBNx{9783642400469}


\bibitem[Csardi et~al\mbox{.}(2006)]%
        {csardi2006igraph}
\bibfield{author}{\bibinfo{person}{G. Csardi}, \bibinfo{person}{T. Nepusz}, {et~al\mbox{.}}} \bibinfo{year}{2006}\natexlab{}.
\newblock \showarticletitle{The igraph software package for complex network research}.
\newblock \bibinfo{journal}{\emph{InterJournal, complex systems}} \bibinfo{volume}{1695}, \bibinfo{number}{5} (\bibinfo{year}{2006}), \bibinfo{pages}{1--9}.
\newblock


\bibitem[Fazlali et~al\mbox{.}(2017)]%
        {com-fazlali17}
\bibfield{author}{\bibinfo{person}{M. Fazlali}, \bibinfo{person}{E. Moradi}, {and} \bibinfo{person}{H. Malazi}.} \bibinfo{year}{2017}\natexlab{}.
\newblock \showarticletitle{{Adaptive parallel Louvain community detection on a multicore platform}}.
\newblock \bibinfo{journal}{\emph{Microprocessors and microsystems}}  \bibinfo{volume}{54} (\bibinfo{date}{Oct} \bibinfo{year}{2017}), \bibinfo{pages}{26--34}.
\newblock


\bibitem[Gach and Hao(2014)]%
        {com-gach14}
\bibfield{author}{\bibinfo{person}{O. Gach} {and} \bibinfo{person}{J. Hao}.} \bibinfo{year}{2014}\natexlab{}.
\newblock \showarticletitle{Improving the Louvain algorithm for community detection with modularity maximization}. In \bibinfo{booktitle}{\emph{Artificial Evolution: 11th International Conference, Evolution Artificielle, EA , Bordeaux, France, October 21-23, . Revised Selected Papers 11}}. Springer, \bibinfo{publisher}{Springer}, \bibinfo{address}{Bordeaux, France}, \bibinfo{pages}{145--156}.
\newblock


\bibitem[Ghosh et~al\mbox{.}(2019)]%
        {com-ghosh19}
\bibfield{author}{\bibinfo{person}{S. Ghosh}, \bibinfo{person}{M. Halappanavar}, \bibinfo{person}{A. Tumeo}, {and} \bibinfo{person}{A. Kalyanarainan}.} \bibinfo{year}{2019}\natexlab{}.
\newblock \showarticletitle{{Scaling and quality of modularity optimization methods for graph clustering}}. In \bibinfo{booktitle}{\emph{IEEE High Performance Extreme Computing Conference (HPEC)}}. \bibinfo{publisher}{IEEE}, \bibinfo{pages}{1--6}.
\newblock
\showISBNx{978-1-7281-5020-8}


\bibitem[Ghosh et~al\mbox{.}(2018)]%
        {com-ghosh18}
\bibfield{author}{\bibinfo{person}{S. Ghosh}, \bibinfo{person}{M. Halappanavar}, \bibinfo{person}{A. Tumeo}, \bibinfo{person}{A. Kalyanaraman}, \bibinfo{person}{H. Lu}, \bibinfo{person}{D. Chavarria-Miranda}, \bibinfo{person}{A. Khan}, {and} \bibinfo{person}{A. Gebremedhin}.} \bibinfo{year}{2018}\natexlab{}.
\newblock \showarticletitle{Distributed louvain algorithm for graph community detection}. In \bibinfo{booktitle}{\emph{IEEE International Parallel and Distributed Processing Symposium (IPDPS)}}. \bibinfo{address}{Vancouver, British Columbia, Canada}, \bibinfo{pages}{885--895}.
\newblock


\bibitem[Gregory(2010)]%
        {com-gregory10}
\bibfield{author}{\bibinfo{person}{S. Gregory}.} \bibinfo{year}{2010}\natexlab{}.
\newblock \showarticletitle{{Finding overlapping communities in networks by label propagation}}.
\newblock \bibinfo{journal}{\emph{New Journal of Physics}}  \bibinfo{volume}{12} (\bibinfo{date}{10} \bibinfo{year}{2010}), \bibinfo{pages}{103018}.
\newblock
Issue 10.


\bibitem[Halappanavar et~al\mbox{.}(2017)]%
        {com-halappanavar17}
\bibfield{author}{\bibinfo{person}{M. Halappanavar}, \bibinfo{person}{H. Lu}, \bibinfo{person}{A. Kalyanaraman}, {and} \bibinfo{person}{A. Tumeo}.} \bibinfo{year}{2017}\natexlab{}.
\newblock \showarticletitle{{Scalable static and dynamic community detection using Grappolo}}. In \bibinfo{booktitle}{\emph{IEEE High Performance Extreme Computing Conference (HPEC)}}. \bibinfo{publisher}{IEEE}, \bibinfo{address}{Waltham, MA USA}, \bibinfo{pages}{1--6}.
\newblock
\showISBNx{978-1-5386-3472-1}


\bibitem[Hu et~al\mbox{.}({[n.\,d.]})]%
        {huparleiden}
\bibfield{author}{\bibinfo{person}{Yongmin Hu}, \bibinfo{person}{Jing Wang}, \bibinfo{person}{Cheng Zhao}, \bibinfo{person}{Yibo Liu}, \bibinfo{person}{Cheng Chen}, \bibinfo{person}{Xiaoliang Cong}, {and} \bibinfo{person}{Chao Li}.} \bibinfo{year}{[n.\,d.]}\natexlab{}.
\newblock \showarticletitle{ParLeiden: Boosting Parallelism of Distributed Leiden Algorithm on Large-scale Graphs}.
\newblock  (\bibinfo{year}{[n.\,d.]}).
\newblock


\bibitem[Kang et~al\mbox{.}(2023)]%
        {kang2023cugraph}
\bibfield{author}{\bibinfo{person}{S. Kang}, \bibinfo{person}{C. Hastings}, \bibinfo{person}{J. Eaton}, {and} \bibinfo{person}{B. Rees}.} \bibinfo{year}{2023}\natexlab{}.
\newblock \showarticletitle{cuGraph C++ primitives: vertex/edge-centric building blocks for parallel graph computing}. In \bibinfo{booktitle}{\emph{IEEE International Parallel and Distributed Processing Symposium Workshops}}. \bibinfo{pages}{226--229}.
\newblock


\bibitem[Kolodziej et~al\mbox{.}(2019)]%
        {suite19}
\bibfield{author}{\bibinfo{person}{S. Kolodziej}, \bibinfo{person}{M. Aznaveh}, \bibinfo{person}{M. Bullock}, \bibinfo{person}{J. David}, \bibinfo{person}{T. Davis}, \bibinfo{person}{M. Henderson}, \bibinfo{person}{Y. Hu}, {and} \bibinfo{person}{R. Sandstrom}.} \bibinfo{year}{2019}\natexlab{}.
\newblock \showarticletitle{{The SuiteSparse matrix collection website interface}}.
\newblock \bibinfo{journal}{\emph{The Journal of Open Source Software}} \bibinfo{volume}{4}, \bibinfo{number}{35} (\bibinfo{date}{Mar} \bibinfo{year}{2019}), \bibinfo{pages}{1244}.
\newblock


\bibitem[Lancichinetti and Fortunato(2009)]%
        {com-lancichinetti09}
\bibfield{author}{\bibinfo{person}{A. Lancichinetti} {and} \bibinfo{person}{S. Fortunato}.} \bibinfo{year}{2009}\natexlab{}.
\newblock \showarticletitle{Community detection algorithms: a comparative analysis.}
\newblock \bibinfo{journal}{\emph{Physical Review. E, Statistical, Nonlinear, and Soft Matter Physics}} \bibinfo{volume}{80}, \bibinfo{number}{5 Pt 2} (\bibinfo{date}{Nov} \bibinfo{year}{2009}), \bibinfo{pages}{056117}.
\newblock


\bibitem[Leskovec(2021)]%
        {com-leskovec21}
\bibfield{author}{\bibinfo{person}{J. Leskovec}.} \bibinfo{year}{2021}\natexlab{}.
\newblock \bibinfo{title}{{CS224W: Machine Learning with Graphs | 2021 | Lecture 13.3 - Louvain Algorithm}}.
\newblock
\newblock
\urldef\tempurl%
\url{https://www.youtube.com/watch?v=0zuiLBOIcsw}
\showURL{%
\tempurl}


\bibitem[Lu et~al\mbox{.}(2015)]%
        {com-lu15}
\bibfield{author}{\bibinfo{person}{H. Lu}, \bibinfo{person}{M. Halappanavar}, {and} \bibinfo{person}{A. Kalyanaraman}.} \bibinfo{year}{2015}\natexlab{}.
\newblock \showarticletitle{{Parallel heuristics for scalable community detection}}.
\newblock \bibinfo{journal}{\emph{Parallel computing}}  \bibinfo{volume}{47} (\bibinfo{date}{Aug} \bibinfo{year}{2015}), \bibinfo{pages}{19--37}.
\newblock


\bibitem[Mohammadi et~al\mbox{.}(2020)]%
        {com-mohammadi20}
\bibfield{author}{\bibinfo{person}{M. Mohammadi}, \bibinfo{person}{M. Fazlali}, {and} \bibinfo{person}{M. Hosseinzadeh}.} \bibinfo{year}{2020}\natexlab{}.
\newblock \showarticletitle{{Accelerating Louvain community detection algorithm on graphic processing unit}}.
\newblock \bibinfo{journal}{\emph{The Journal of supercomputing}} (\bibinfo{date}{Nov} \bibinfo{year}{2020}).
\newblock


\bibitem[Naim et~al\mbox{.}(2017)]%
        {com-naim17}
\bibfield{author}{\bibinfo{person}{M. Naim}, \bibinfo{person}{F. Manne}, \bibinfo{person}{M. Halappanavar}, {and} \bibinfo{person}{A. Tumeo}.} \bibinfo{year}{2017}\natexlab{}.
\newblock \showarticletitle{{Community detection on the GPU}}. In \bibinfo{booktitle}{\emph{IEEE International Parallel and Distributed Processing Symposium (IPDPS)}}. \bibinfo{publisher}{IEEE}, \bibinfo{address}{Orlando, Florida, USA}, \bibinfo{pages}{625--634}.
\newblock
\showISBNx{978-1-5386-3914-6}


\bibitem[Newman(2006)]%
        {com-newman06}
\bibfield{author}{\bibinfo{person}{M. Newman}.} \bibinfo{year}{2006}\natexlab{}.
\newblock \showarticletitle{{Finding community structure in networks using the eigenvectors of matrices}}.
\newblock \bibinfo{journal}{\emph{Physical review E}} \bibinfo{volume}{74}, \bibinfo{number}{3} (\bibinfo{year}{2006}), \bibinfo{pages}{036104}.
\newblock


\bibitem[Nguyen({[n.\,d.]})]%
        {nguyenleiden}
\bibfield{author}{\bibinfo{person}{Fabian Nguyen}.} \bibinfo{year}{[n.\,d.]}\natexlab{}.
\newblock \emph{\bibinfo{title}{Leiden-Based Parallel Community Detection}}.
\newblock Bachelor's Thesis. \bibinfo{school}{Karlsruhe Institute of Technology, 2021 (zitiert auf S. 31)}.
\newblock


\bibitem[Rotta and Noack(2011)]%
        {com-rotta11}
\bibfield{author}{\bibinfo{person}{R. Rotta} {and} \bibinfo{person}{A. Noack}.} \bibinfo{year}{2011}\natexlab{}.
\newblock \showarticletitle{Multilevel local search algorithms for modularity clustering}.
\newblock \bibinfo{journal}{\emph{Journal of Experimental Algorithmics (JEA)}}  \bibinfo{volume}{16} (\bibinfo{year}{2011}), \bibinfo{pages}{2--1}.
\newblock


\bibitem[Ryu and Kim(2016)]%
        {com-ryu16}
\bibfield{author}{\bibinfo{person}{S. Ryu} {and} \bibinfo{person}{D. Kim}.} \bibinfo{year}{2016}\natexlab{}.
\newblock \showarticletitle{{Quick community detection of big graph data using modified louvain algorithm}}. In \bibinfo{booktitle}{\emph{IEEE 18th International Conference on High Performance Computing and Communications (HPCC)}}. \bibinfo{publisher}{IEEE}, \bibinfo{address}{Sydney, NSW}, \bibinfo{pages}{1442--1445}.
\newblock
\showISBNx{978-1-5090-4297-5}


\bibitem[Sahu(2023a)]%
        {report}
\bibfield{author}{\bibinfo{person}{S. Sahu}.} \bibinfo{year}{2023}\natexlab{a}.
\newblock \showarticletitle{GVE-Leiden: Fast Leiden Algorithm for Community Detection in Shared Memory Setting}.
\newblock \bibinfo{journal}{\emph{arXiv preprint arXiv:2312.13936}} (\bibinfo{year}{2023}).
\newblock


\bibitem[Sahu(2023b)]%
        {sahu2023gvelouvain}
\bibfield{author}{\bibinfo{person}{Subhajit Sahu}.} \bibinfo{year}{2023}\natexlab{b}.
\newblock \showarticletitle{GVE-Louvain: Fast Louvain Algorithm for Community Detection in Shared Memory Setting}.
\newblock \bibinfo{journal}{\emph{arXiv preprint arXiv:2312.04876}} (\bibinfo{year}{2023}).
\newblock


\bibitem[Sattar and Arifuzzaman(2019)]%
        {com-sattar19}
\bibfield{author}{\bibinfo{person}{N. Sattar} {and} \bibinfo{person}{S. Arifuzzaman}.} \bibinfo{year}{2019}\natexlab{}.
\newblock \showarticletitle{Overcoming MPI Communication Overhead for Distributed Community Detection}. In \bibinfo{booktitle}{\emph{Software Challenges to Exascale Computing}}, \bibfield{editor}{\bibinfo{person}{A.~Majumdar} {and} \bibinfo{person}{R.~Arora}} (Eds.). \bibinfo{publisher}{Springer Singapore}, \bibinfo{address}{Singapore}, \bibinfo{pages}{77--90}.
\newblock
\showISBNx{978-981-13-7729-7}


\bibitem[Shi et~al\mbox{.}(2021)]%
        {com-shi21}
\bibfield{author}{\bibinfo{person}{J. Shi}, \bibinfo{person}{L. Dhulipala}, \bibinfo{person}{D. Eisenstat}, \bibinfo{person}{J. {\L}{\k{a}}cki}, {and} \bibinfo{person}{V. Mirrokni}.} \bibinfo{year}{2021}\natexlab{}.
\newblock \bibinfo{title}{Scalable community detection via parallel correlation clustering}.
\newblock
\newblock


\bibitem[Staudt et~al\mbox{.}(2016)]%
        {staudt2016networkit}
\bibfield{author}{\bibinfo{person}{C.L. Staudt}, \bibinfo{person}{A. Sazonovs}, {and} \bibinfo{person}{H. Meyerhenke}.} \bibinfo{year}{2016}\natexlab{}.
\newblock \showarticletitle{NetworKit: A tool suite for large-scale complex network analysis}.
\newblock \bibinfo{journal}{\emph{Network Science}} \bibinfo{volume}{4}, \bibinfo{number}{4} (\bibinfo{year}{2016}), \bibinfo{pages}{508--530}.
\newblock


\bibitem[Traag(2015)]%
        {com-traag15}
\bibfield{author}{\bibinfo{person}{V. Traag}.} \bibinfo{year}{2015}\natexlab{}.
\newblock \showarticletitle{Faster unfolding of communities: Speeding up the Louvain algorithm}.
\newblock \bibinfo{journal}{\emph{Physical Review E}} \bibinfo{volume}{92}, \bibinfo{number}{3} (\bibinfo{year}{2015}), \bibinfo{pages}{032801}.
\newblock


\bibitem[Traag(2024)]%
        {traag2024leiden}
\bibfield{author}{\bibinfo{person}{V. Traag}.} \bibinfo{year}{2024}\natexlab{}.
\newblock \showarticletitle{Personal communication}.
\newblock  (\bibinfo{year}{2024}).
\newblock


\bibitem[Traag et~al\mbox{.}(2011)]%
        {com-traag11}
\bibfield{author}{\bibinfo{person}{V. Traag}, \bibinfo{person}{P. Dooren}, {and} \bibinfo{person}{Y. Nesterov}.} \bibinfo{year}{2011}\natexlab{}.
\newblock \showarticletitle{{Narrow scope for resolution-limit-free community detection}}.
\newblock \bibinfo{journal}{\emph{Physical Review E}} \bibinfo{volume}{84}, \bibinfo{number}{1} (\bibinfo{year}{2011}), \bibinfo{pages}{016114}.
\newblock


\bibitem[Traag et~al\mbox{.}(2019)]%
        {com-traag19}
\bibfield{author}{\bibinfo{person}{V. Traag}, \bibinfo{person}{L. Waltman}, {and} \bibinfo{person}{N. Eck}.} \bibinfo{year}{2019}\natexlab{}.
\newblock \showarticletitle{{From Louvain to Leiden: guaranteeing well-connected communities.}}
\newblock \bibinfo{journal}{\emph{Scientific Reports}} \bibinfo{volume}{9}, \bibinfo{number}{1} (\bibinfo{date}{Mar} \bibinfo{year}{2019}), \bibinfo{pages}{5233}.
\newblock


\bibitem[Verweij({[n.\,d.]})]%
        {verweijfaster}
\bibfield{author}{\bibinfo{person}{Geerten Verweij}.} \bibinfo{year}{[n.\,d.]}\natexlab{}.
\newblock \emph{\bibinfo{title}{Faster Community Detection Without Loss of Quality: Parallelizing the Leiden Algorithm}}.
\newblock Master's Thesis. \bibinfo{school}{Leiden University, 2020}.
\newblock


\bibitem[Waltman and Eck(2013)]%
        {com-waltman13}
\bibfield{author}{\bibinfo{person}{L. Waltman} {and} \bibinfo{person}{N. Eck}.} \bibinfo{year}{2013}\natexlab{}.
\newblock \showarticletitle{A smart local moving algorithm for large-scale modularity-based community detection}.
\newblock \bibinfo{journal}{\emph{The European physical journal B}} \bibinfo{volume}{86}, \bibinfo{number}{11} (\bibinfo{year}{2013}), \bibinfo{pages}{1--14}.
\newblock


\bibitem[Wickramaarachchi et~al\mbox{.}(2014)]%
        {com-wickramaarachchi14}
\bibfield{author}{\bibinfo{person}{C. Wickramaarachchi}, \bibinfo{person}{M. Frincu}, \bibinfo{person}{P. Small}, {and} \bibinfo{person}{V. Prasanna}.} \bibinfo{year}{2014}\natexlab{}.
\newblock \showarticletitle{Fast parallel algorithm for unfolding of communities in large graphs}. In \bibinfo{booktitle}{\emph{IEEE High Performance Extreme Computing Conference (HPEC)}}. IEEE, \bibinfo{publisher}{IEEE}, \bibinfo{address}{Waltham, MA USA}, \bibinfo{pages}{1--6}.
\newblock


\bibitem[You et~al\mbox{.}(2022)]%
        {com-you22}
\bibfield{author}{\bibinfo{person}{Y. You}, \bibinfo{person}{L. Ren}, \bibinfo{person}{Z. Zhang}, \bibinfo{person}{K. Zhang}, {and} \bibinfo{person}{J. Huang}.} \bibinfo{year}{2022}\natexlab{}.
\newblock \showarticletitle{Research on improvement of Louvain community detection algorithm}. In \bibinfo{booktitle}{\emph{2nd International Conference on Artificial Intelligence, Automation, and High-Performance Computing (AIAHPC )}}, Vol.~\bibinfo{volume}{12348}. \bibinfo{publisher}{SPIE}, \bibinfo{address}{Zhuhai, China}, \bibinfo{pages}{527--531}.
\newblock


\bibitem[Zhang et~al\mbox{.}(2021)]%
        {com-zhang21}
\bibfield{author}{\bibinfo{person}{J. Zhang}, \bibinfo{person}{J. Fei}, \bibinfo{person}{X. Song}, {and} \bibinfo{person}{J. Feng}.} \bibinfo{year}{2021}\natexlab{}.
\newblock \showarticletitle{An improved Louvain algorithm for community detection}.
\newblock \bibinfo{journal}{\emph{Mathematical Problems in Engineering}}  \bibinfo{volume}{2021} (\bibinfo{year}{2021}), \bibinfo{pages}{1--14}.
\newblock


\end{thebibliography}

\clearpage
\appendix
\section{Appendix}

\subsection{Finding disconnected communities}

We now outline our parallel algorithm for identifying disconnected communities, given the original graph and the community membership of each vertex. The core concept involves determining the size of each community, selecting a vertex from each community, traversing within the community from that vertex (avoiding adjacent communities), and marking a community as disconnected if all its vertices cannot be reached. We explore four distinct approaches, differing in the use of parallel Depth-First Search (DFS) or Breadth-First Search (BFS) and whether per-thread or shared \textit{visited} flags are employed. If shared visited flags are used, each thread scans all vertices but processes only its assigned community based on the community ID. Our findings suggest that utilizing parallel BFS traversal with a shared flag vector yields the fastest results. As this is not a heuristic algorithm, all approaches produce identical outcomes. Algorithm \ref{alg:disconnected} illustrates the pseudocode for this approach. Here, the \texttt{disconnectedCommunities()} function takes the input graph $G$ and the community membership $C$ as input, and it returns the disconnected flag $D$ for each community.

We now explain Algorithm \ref{alg:disconnected} in detail. First, in line \ref{alg:disconnected--init}, the disconnected community flag $D$, and the visited vertices flags $vis$ are initialized. In line \ref{alg:disconnected--sizes}, the size of each community $S$ is obtained in parallel using the \texttt{communitySizes()} function. Subsequently, each thread processes each vertex $i$ in the graph $G$ in parallel (lines \ref{alg:disconnected--loop-begin}-\ref{alg:disconnected--loop-end}). In line \ref{alg:disconnected--unreached}, the community membership of $i$ ($c$) is determined, and the count of vertices reached from $i$ is initialized to $0$. If community $c$ is empty or not in the work-list of the current thread $work_t$, the thread proceeds to the next iteration (line \ref{alg:disconnected--work}). If however the community $c$ is non-empty and in the work-list of the current thread $work_t$, BFS is performed from vertex $i$ to explore vertices in the same community, using lambda functions $f_{if}$ to conditionally perform BFS to vertex $j$ if it belongs to the same community, and $f_{do}$ to update the count of $reached$ vertices after each vertex is visited during BFS (line \ref{alg:disconnected--bfs}). If the number of vertices $reached$ during BFS is less than the community size $S[c]$, the community $c$ is marked as disconnected (line \ref{alg:disconnected--mark}). Finally, the size of the community $S[c]$ is updated to $0$, indicating that the community has been processed (line \ref{alg:disconnected--processed}). Note that the work-list $work_t$ for each thread with ID $t$, is defined as a set containing communities $[t\chi,\ t(\chi+1))\ \cup\ [T\chi + t\chi,\ T\chi + t(\chi+1))\ \cup\ \ldots$, where $\chi$ is the chunk size, and $T$ is the number of threads. We use a chunk size of $\chi = 1024$.

\begin{algorithm}[hbtp]
\caption{Finding disconnected communities in parallel.}
\label{alg:disconnected}
\begin{algorithmic}[1]
\Require{$G$: Input graph}
\Require{$C$: Community membership of each vertex}
\Ensure{$D$: Disconnected flag for each community}
\Ensure{$S$: Size of each community}
\Ensure{$f_{if}$: Perform BFS to vertex $j$ if condition satisfied}
\Ensure{$f_{do}$: Perform operation after each vertex is visited}
\Ensure{$reached$: Number of vertices reachable from $i$ to $i$'s community}
\Ensure{$work_t$: Work-list of current thread}

\Statex

\Function{disconnectedCommunities}{$G, C$} \label{alg:disconnected--begin}
  \State $D \gets \{\}$ \textbf{;} $vis \gets \{\}$ \label{alg:disconnected--init}
  \State $S \gets communitySizes(G, C)$ \label{alg:disconnected--sizes}
  \ForAll{\textbf{threads in parallel}} \label{alg:disconnected--threads-begin}
    \ForAll{$i \in V$} \label{alg:disconnected--loop-begin}
      \State $c \gets C[i]$ \textbf{;} $reached \gets 0$ \label{alg:disconnected--unreached}
      \State $\rhd$ Skip if community $c$ is empty, or
      \State $\rhd$ does not belong to work-list of current thread.
      \If{$S[c] = 0$ \textbf{or} $c \notin work_t$} \textbf{continue} \label{alg:disconnected--work}
      \EndIf
      \State $f_{if} \gets (j) \implies C[j] = c$
      \State $f_{do} \gets (j) \implies reached \gets reached + 1$
      \State $bfsVisitForEach(vis, G, i, f_{if}, f_{do})$ \label{alg:disconnected--bfs}
      \If{$reached < S[c]$} $D[c] \gets 1$ \label{alg:disconnected--mark}
      \EndIf
      \State $S[c] \gets 0$ \label{alg:disconnected--processed}
    \EndFor \label{alg:disconnected--loop-end}
  \EndFor \label{alg:disconnected--threads-end}
  \Return{$D$}
\EndFunction \label{alg:disconnected--end}
\end{algorithmic}
\end{algorithm}

\subsection{Indirect Comparison with State-of-the-art Leiden Implementations}
\label{sec:comparison-indirect}

Finally, we conduct an indirect comparison of the performance of our multicore implementation of the Leiden algorithm with other similar state-of-the-art implementations, as listed in Table \ref{tab:compare}. Please consider the reported speedups as approximate. Hu et al. \cite{huparleiden} introduce ParLeiden, a parallel Leiden implementation for distributed environments, which uses thread locks and efficient buffers, to resolve community joining conflicts and reduce communication overheads. They refer to their single node version of ParLeiden as ParLeiden-S, and their distributed version as ParLeiden-D. On a cluster with $8$ nodes, with each node being equipped with a $48$ core CPU, Hu et al. observe a speedup of $12.3\times$, $9.9\times$, and $1.32\times$ for ParLeiden-S, ParLeiden-D, and a baseline Leiden implemented on KatanaGraph, on the \textit{com-LiveJournal} graph, with respect to original Leiden \cite{com-traag19} (refer to Table 2 in their paper \cite{huparleiden}). In contrast, on the same graph, we observe a speedup of $219\times$ relative to original Leiden. Consequently, GVE-Leiden outperforms ParLeiden-S, ParLeiden-D, and KatanaGraph Leiden by approximately $18\times$, $22\times$, and $166\times$ respectively\ignore{ (all achieved without GPU)}.

\end{document}